\documentclass[10pt,letterpaper,english,twoside,onecolumn]{article}
\usepackage{amsmath}
\usepackage{amssymb}
\usepackage{graphicx}
\usepackage{color}
\usepackage{bm}
\usepackage{ulem} 
\usepackage[outercaption]{sidecap}
\usepackage{geometry}
\def\vts{v_{{\rm t}s}}
\def\vte{v_{{\rm te}}}
\def\vti{v_{{\rm ti}}}
\def\vtE{v_{{\rm tE}}}

\def\rhoLs{\rho_{{\rm c}s}}
\def\rhoLe{\rho_{{\rm ce}}}
\def\rhoLi{\rho_{{\rm ci}}}
\def\rhoLE{\rho_{{\rm cE}}}

\def\omLsO{\omega_{{\rm c}0s}}
\def\omLs{\omega_{{\rm c}s}}

\def\omLi{\omega_{{\rm ci}}}

\def\omegad{\omega_{\rm d}}
\def\omegads{\omega_{{\rm d}s}}
\def\omegadi{\omega_{\rm di}}

\def\vA{v_{\rm A}}
\def\vAO{v_{{\rm A0}}}
\def\omA{\omega_{{\rm A}}}
\def\omAO{\omega_{{\rm A0}}}

\def\me{m_{\rm e}}
\def\mi{m_{\rm i}}

\def\Ts{\mathcal{T}_s}
\def\Te{\mathcal{T}_{\rm e}}
\def\Ti{\mathcal{T}_{\rm i}}
\def\TE{\mathcal{T}_{\rm E}}
\def\taues{\tau_{{\rm e}s}}
\def\tausi{\tau_{s{\rm i}}}
\def\tauis{\tau_{{\rm i}s}}
\def\tauei{\tau_{\rm ei}}
\def\tauiE{\tau_{\rm iE}}
\def\tauEi{\tau_{\rm Ei}}

\def\feqs{f_{0s}}
\def\FM{F_{\rm M}}
\def\FMs{F_{{\rm M}s}}
\def\FMi{F_{\rm Mi}}
\def\nsO{n_{0s}}
\def\nEO{n_{0{\rm E}}}
\def\niO{n_{0{\rm i}}}

\def\F{\mathcal{F}}

\def\L{\mathcal{L}}
\def\T{\mathcal{T}}
\def\O{\mathcal{O}}
\def\enr{\mathcal{E}}

\def\dphi{\delta\phi}
\def\dpsi{\delta\psi}
\def\dBp{\delta B_\parallel}

\def\eps{\varepsilon}

\def\sgnVp{\hat{\sigma}}
\def\delds{\delta_{{\rm d}s}}
\def\dGds{\delta G_{{\rm d}s}}

\def\dEe{\delta E_{\rm e}}
\def\dPsie{\delta\Psi_{\rm e}}
\def\dUe{\delta U_{\rm e}}
\def\dCe{\delta C_{\rm e}}

\def\ks{\hat{k}_s}
\def\ksO{\hat{k}_{0s}}

\def\kiO{\hat{k}_{0{\rm i}}}

\def\kEO{\hat{k}_{0{\rm E}}}

\def\acriti{\alpha_{\rm crit}^{\rm (I)}}
\def\acritii{\alpha_{\rm crit}^{\rm (II)}}

\definecolor{gray}{rgb}{0.5,0.5,0.5}
\definecolor{dred}{rgb}{0.5,0.0,0.0}
\definecolor{dgreen}{rgb}{0.0,0.5,0.0}
\definecolor{dblue}{rgb}{0.0,0.0,0.5}

\begin{document}
\title{AWECS: A Linear Gyrokinetic $\delta f$ Particle-in-Cell \\
Simulation Code for the Study of Alfv\'{e}nic Instabilities \\
in High-$\beta$ Tokamak Plasmas}

\author{Andreas Bierwage$^{1,}$\thanks{abierwag@uci.edu}\, and Liu Chen$^{1,2}$ \\
$^1$ {\it\normalsize Department of Physics and Astronomy, University of California, Irvine, CA 92697, U.S.A.} \\
$^2$ {\it\normalsize Institute for Fusion Theory and Simulation, Zhejiang University, Hangzhou, China}}

\date{\today}

\maketitle

\begin{abstract}
A 1-D linear gyrokinetic code called \textsc{awecs} is developed to study the kinetic excitation of Alfv\'{e}nic instabilities in a high-$\beta$ tokamak plasma, with $\beta$ being the ratio of thermal to magnetic pressure. It is designed to describe physics associated with a broad range of frequencies and wavelengths. For example, \textsc{awecs} is capable of simulating kinetic ballooning modes, Alfv\'{e}nic ion-temperature-gradient-driven modes, as well as Alfv\'{e}n instabilities due to energetic particles. In addition, \textsc{awecs} may be used to study drift-Alfv\'{e}n instabilities in the low-$\beta$ regime. Here, the layout of the code and the numerical methods used are described. \textsc{awecs} is benchmarked against other codes and a convergence study is carried out.
\end{abstract}



\section{Introduction}
\label{sec:intro}

In a tokamak, nested closed toroidal magnetic surfaces are used to confine a high-temperature plasma consisting mainly of ionized Deuterium. Such magnetized plasmas are known to support various kinds of magnetohydrodynamic (MHD) shear Alfv\'{e}n waves (SAW), the properties of which are determined by the geometry of the magnetic flux surfaces; in particular, the field line curvature and magnetic shear. Resonant and non-resonant interactions between SAWs and plasma particles can lead to excitations of SAW instabilities. These instabilities may, in turn, affect particle confinement. In order to optimize the tokamak geometry and operating conditions for thermonuclear fusion applications, a thorough understanding of SAW physics is crucial. For a review of SAW observations and comparison with theory see, e.g., Ref.~\cite{Wong99}.

In order to investigate the linear instability of SAWs, a linear gyrokinetic particle-in-cell (PIC) simulation code, called \textsc{awecs}, is developed. The model equations describe the dynamics local to a field-aligned flux-tube using the so-called ballooning formalism. The equations are valid for a broad range of frequencies and wavelengths, with focus on temperature- and pressure-gradient-driven instabilities, while ignoring modes driven by the gradient of the parallel plasma current. For instance, \textsc{awecs} allows to study electrostatic and Alfv\'{e}nic ion-temperature-gradient modes (ESITG and AITG) \cite{Romanelli89, Zonca98}, kinetic ballooning modes (KBM) \cite{Tsai93}, $\beta$-induced Alfv\'{e}n eigenmodes (BAE) \cite{Chu92}, toroidicity-induced Alfv\'{e}n eigenmodes (TAE) \cite{Cheng85}, $\alpha$-induced toroidal Alfv\'{e}n eigenmodes ($\alpha$TAE) \cite{Hu04}, as well as energetic particle modes (EPM) \cite{Chen94}. In addition, \textsc{awecs} may be used to study drift-Alfv\'{e}n instabilities in the low-$\beta$ regime \cite{Mikhailovskii}. Here, $\beta = 2\mu_0 P / B_0^2$ is the ratio of thermal to magnetic pressure, and $\alpha = -q^2 R_0 {\rm d}\beta/{\rm d}r$ is the normalized pressure gradient, with $q$ being the safety factor (a measure for the field line pitch), $R_0$ the major radius of the torus, $r$ the minor radial coordinate, and $B_0$ the field strength at the magnetic axis.

This paper is organized as follows. In Section~\ref{sec:model}, we describe the physical model and, in Section~\ref{sec:num}, the numerical methods used to solve the equations. In Sections~\ref{sec:bench_esitg}--\ref{sec:bench_atae}, \textsc{awecs} is benchmarked against other codes, followed by a convergence study in Section~\ref{sec:accurate_converge}. Concluding remarks and discussions are given in Section~\ref{sec:conclusion}.

\section{Model}
\label{sec:model}

In this section, we describe the physical model used. After providing an overview of assumptions made in the derivation, we describe the equilibrium model, followed by the gyrokinetic equation and the electromagnetic field equations. Then the equations are normalized and cast into a form suitable for numerical solution as an initial-value problem. For convenience, in the first part of this section, all time-dependent variables are Laplace-transformed ($\partial_t \equiv \partial/\partial t \rightarrow -i\omega$).

\subsection{Assumptions and formal ordering}
\label{sec:model_approx}

We employ the linear gyrokinetic field equations derived by Zonca \& Chen \cite{Zonca06}. A reduction similar to that described in Ref.~\cite{Zonca06} is applied, except that finite-Larmor-radius (FLR) corrections for thermal ions are retained in the present paper. The model is valid for Alfv\'{e}nic instabilities in a wide range of frequencies $\omega$ and wave numbers $k$, provided that
\begin{equation}
{\rm (i)}:\; \omega \ll k_\parallel \vte, \quad k_\perp\rhoLe \ll 1, \qquad
{\rm (ii)}:\; \omega \ll \omLi, \qquad
{\rm (iii)}:\; \omega \lesssim \omA;
\label{eq:model_approx}
\end{equation}

\noindent where $\omLs = e_s B/m_s$ is the cyclotron frequency and $\rhoLs = v_\perp/\omLs$ the Larmor radius for particle species $s$ ($s = {\rm e}$ for electrons, $s = {\rm i}$ for thermal ions, and $s = {\rm E}$ for energetic ions). Here, $k_\parallel$ is a short-hand notation for the typical value of $({\bm B}/B)\cdot{\bm\nabla} = \partial/\partial l$ in regions where the field perturbation has a significant amplitude. Correspondingly, $k_\perp$ measures the wave number perpendicular to the equilibrium magnetic field ${\bm B}$. The thermal velocity $\vts$ is defined as $T_s = m_s\vts^2$. The restrictions in Eq.~(\ref{eq:model_approx}) mean that (i) we consider wave dynamics which are adiabatic with respect to electron dynamics and neglect electron-FLR effects (formally: $\me \rightarrow 0^+$), (ii) the magnetic moment is conserved (no cyclotron resonances), and (iii) the interaction with fast magnetosonic waves is negligible. Although, assumption (i), $\omega \ll k_\parallel \vte$, is applicable only to passing electrons, kinetic effects associated with magnetically trapped electrons are ignored at this stage. Assumption (ii) implies that the only relevant ``kinetic effect'' is the kinetic compression of ions along the magnetic field, so the particle dynamics are described in terms of the parallel velocity $v_\parallel$ and the position $l$ along a field line.

In this study, effects associated with the anisotropy of the equilibrium particle distribution $f_0$ are neglected; thus, we let $P = P_\perp \approx P_\parallel$. Furthermore, since we are dealing with low-frequency waves, $\omega/k \ll c$, the displacement current in Amp\`{e}re's law is neglected (${\bm \nabla}\times{\bm B} \approx \mu_0 {\bm j}$). As is typical for tokamaks, the plasma is taken to be sufficiently dense to satisfy the quasi-neutrality condition; i.e., $k_\perp \ll 1/\lambda_{\rm De}$, with $\lambda_{\rm De}$ being the electron Debye length.

The equations are linearized by separating the distribution $\F_s$ (phase-space density) and the electromagnetic fields ${\bm E}_{\rm tot}$ and ${\bm B}_{\rm tot}$ into equilibrium and perturbed components,
\begin{equation}
\F_s = \feqs + \delta f_s, \qquad
{\bm B}_{\rm tot} = {\bm\nabla}\times{\bm A}_{\rm tot} = {\bm B} + \delta{\bm B}, \qquad
{\bm E}_{\rm tot} = \delta{\bm E} = -{\bm\nabla}\delta\phi - \partial_t\delta{\bm A};
\end{equation}

\noindent where we have assumed that the equilibrium is static (${\bm E}_{\rm tot} = \delta{\bm E}$). We use the Coulomb gauge ${\bm\nabla}\cdot{\bm A}_{\rm tot} = 0$. Since we consider micro-scale instabilities, the length scales of equilibrium and perturbed quantities are disparate in the direction perpendicular to the magnetic field; i.e., $k_\perp \gg k_{0\perp}$. Due to the stabilizing influence of magnetic tension, perturbations tend to be aligned with the magnetic field, so that $k_\perp \gg k_\parallel$ and the perturbations have a nearly flute-like structure.

In a tokamak, the toroidal component of the magnetic field is much stronger than the poloidal component, so that $B \approx B_{\rm T}$. In simple toroidal coordinates $(r, \vartheta, \zeta)$ (minor radius, poloidal/azimuthal angle, toroidal angle), $B_{\rm T}$ has the form
\begin{equation}
B_{\rm T} = B_0/\hat{R}, \qquad {\rm with} \quad \hat{R} = R/R_0 = 1 + \eps\cos\vartheta.
\end{equation}

\noindent The typical value for the inverse aspect ratio for the flux surface under consideration, $\eps = r/R_0$, is taken to be of the order $\eps \sim 0.1 \cdots 0.3$, depending on the proximity to the magnetic axis. The scale length of the equilibrium pressure gradient is taken to be of the order $L_p/R_0 \sim 0.1$. Using $\delta \sim L_p/R_0$ as a small expansion parameter, and noting that $k_{0\perp} \sim L_p^{-1}$, we adopt the following formal ordering \cite{Zonca06}:\footnote{In Ref.~\protect\cite{Zonca06}, the ordering of the perpendicular wave number is $k_\perp\rhoLi \sim \O(\delta)$. Here, larger values are allowed and FLR corrections will be included.}
\begin{subequations}
\begin{align}
&\omega / \omA \sim \O(\delta^{1/2});
\label{eq:model_order_w}
\\
&k_{0\parallel}\rhoLi \sim k_{0\perp} \rhoLi \sim \O(\delta), \qquad
k_\parallel \rhoLi \sim \O(\delta), \qquad
k_\perp\rhoLi \sim 1;
\label{eq:model_order_k}
\\
&\frac{\nEO}{\niO} \sim \O(\delta^{3/2}), \qquad
\beta_{\rm i} = \frac{2 T_{\rm i}}{\mi\vAO^2} \sim \O(\delta), \qquad
\beta_{\rm E} = \frac{2 T_{\rm E}}{\mi\vAO^2} \frac{\nEO}{\niO} \sim \O(\delta^{3/2});
\label{eq:model_order_p}
\end{align}
\label{eq:model_order}
\end{subequations}

\noindent where $\vAO = B_0/\sqrt{\mu_0\mi\niO}$ is the Alfv\'{e}n velocity. Furthermore, using the disparity between electron and ion mass, $\me/\mi \sim \O(\delta^3)$, and assuming $T_{\rm e}/T_{\rm i} \sim \O(1)$, we have
\begin{equation}
v_{\rm i} / v_{\rm e} \sim j_{0\parallel{\rm i}} / j_{0\parallel{\rm e}} \sim k_\perp \rho_{\rm e}  \sim \O(\delta^{3/2}) \quad \rightarrow \quad j_{0\parallel} \approx j_{0\parallel{\rm e}}.
\end{equation}

\noindent Thus, we may assume that the equilibrium current is carried primarily by electrons. From the above orderings (\ref{eq:model_order_w})--(\ref{eq:model_order_p}) it follows that
\begin{equation}
\frac{\omega}{\omLi} = \frac{\omega}{\omAO} \frac{\vA}{\vti}k_\vartheta\rhoLi \sim \O(\delta^{3/2}). \nonumber
\end{equation}

\noindent The density and $\beta$ orderings (\ref{eq:model_order_p}) imply that
\begin{equation}
T_{\rm i}/T_{\rm E} \sim \O(\delta), \qquad k_\perp\rhoLE \sim \O(\delta^{-1/2}). \nonumber
\end{equation}

\noindent The wave number ordering (\ref{eq:model_order_k}) implies that \cite{Antonsen80}
\begin{equation}
i{\bm k}_\perp\cdot\delta{\bm A} / \delta A_\parallel \sim \O(\delta). \nonumber
\end{equation}

Note that the above constitutes a maximal ordering, chosen to cover a broad range of plasma parameters; more typical values for the normalized wavenumber used in simulations are, for instance, $k_\perp\rhoLi \sim 0.1$ and $k_\perp\rhoLE \sim 1$, which corresponds to $T_{\rm i}/T_{\rm E} \sim \O(\delta^2)$.

\subsection{Equilibrium model}
\label{sec:model_equlib}

The separation of temporal and spatial scales described by Eq.~(\ref{eq:model_order}), in particular, $\omega \ll \omLi$ and $k_0 \sim k_\parallel \ll k_\perp$, allow us to average over the rapid gyromotion and apply an eikonal approximation, which simplifies the problem significantly. The eikonal approximation is facilitated by the so-called ballooning transform \cite{Connor78}. The essential physical effects arising in toroidal geometry are then captured by the so-called $s$-$\alpha$ model \cite{Connor78}.

For the equilibrium distribution, the separation of temporal and spatial scales implies that, at lowest order, $\feqs$ may be taken to be independent of the position along a field line $l$ and the gyrophase $\xi$ \cite{Antonsen80},
\begin{equation}
\partial_l\feqs = \partial_\xi\feqs = 0.
\label{eq:model_equlib_feq1}
\end{equation}

\noindent Since the field lines cover magnetic surfaces ergodically, Eq.~(\ref{eq:model_equlib_feq1}) implies that the equilibrium particle density $\nsO$ is a function of the minor radius $r$ only. The plasma particle distribution is assumed to be an isotropic ($T_\perp = T_\parallel$) equilibrium, and we usually decompose it as $\feqs = \nsO(r) F_{0s}(r,\enr)$. Thermal ions and electrons are taken to obey a Maxwellian velocity distribution
\begin{equation}
\feqs = \nsO(r) \FMs(r,\enr), \qquad \FMs = [2\pi\Ts(r)]^{-3/2} e^{-\enr/\Ts(r)};
\label{eq:model_equlib_feq2}
\end{equation}

\noindent where
\begin{equation}
\enr = (v_\perp^2 + v_\parallel^2)/2, \qquad \Ts = k_{\rm B} T_s / m_s. \nonumber
\end{equation}

\noindent Energetic ions, such as $\alpha$-particles from nuclear fusion reactions and injected beam ions, are generally non-Maxwellian and anisotropic. In the present model, isotropicity is assumed for simplicity, while allowing for a non-Maxwellian distribution (e.g., slowing-down distribution).

Due to axisymmetry, the tokamak equilibrium is effectively a 2-D system, with $\zeta$ being the ignorable coordinate. Taking advantage of the spatial scale separation, an eikonal approximation would allow to decouple the remaining two dimensions to create two simpler 1-D problems; however, its direct application is infeasible because the magnetic shear severely distorts the flux tubes, while all physical quantities must be $2\pi$-periodic in $\vartheta$ and $\zeta$. The periodicity constraint is eliminated by the ballooning transform, which maps a flux tube of the periodic physical system onto a non-periodic covering space, with the new coordinate being the ballooning angle $\theta \in (-\infty, \infty)$. The physical solution is reconstructed from the solutions on the infinite domain by linear superposition. Formally, the ballooning transform in an axisymmetric configuration can be written as \cite{Connor78}
\begin{equation}
\dphi(r, \vartheta) = \frac{1}{2\pi} \sum\limits_{m=-\infty}^\infty e^{-im\vartheta} \int\limits_{-\infty}^\infty{\rm d}\theta\; e^{-i\theta(nq-m)} \delta\hat{\phi}(r, \theta)
= \sum\limits_{p=-\infty}^{\infty} \delta\hat{\phi}(r,\vartheta + 2\pi p);
\end{equation}

\noindent where $m$ is the poloidal and $n$ the toroidal mode number. The transform is valid for short-wavelength modes ($nq \gg 1$). The dependence on $r$ is then separated as
\begin{equation}
\delta\hat{\phi}(r,\theta) = \delta\hat{\Phi}(\theta) W(r) e^{iS}, \nonumber
\end{equation}

\noindent where $S$ is the eikonal (rapidly varying with $r$) and $W$ the slowly varying radial envelope.

In order to proceed further, one has to construct a local model for the MHD equilibrium. A popular choice is the $s$-$\alpha$ model by Connor, Hastie and Taylor (CHT) \cite{Connor78}, where we may write $\delta\hat{\phi}(r,\theta) = \delta\hat{\Phi}(\theta) e^{i k_r r}$. In the simple case of a cylindrical tokamak, the radial wave number is $k_r = -k_\vartheta s(\theta - \theta_k)$, where $s = (r/q) {\rm d}q/{\rm d}r$ is the global magnetic shear, $\theta_k \propto k_{0\perp}$ describes the slowly-varying radial WKB envelope, and $k_\vartheta = nq/r$ is the poloidal wave number. In toroidal geometry, the thermal pressure modifies the flux surfaces and, thus, imposes a poloidal modulation on the magnetic shear. The CHT $s$-$\alpha$ equilibrium model retains the lowest-order effect by assuming that the toroidal magnetic flux surfaces maintain a circular poloidal cross-section while undergoing an outward shift (Shafranov shift). In this case, the gradient and Laplacian operators applied to perturbed fields, and the magnetic curvature drift, have the following form:
\begin{subequations}
\begin{align}
&-\frac{\partial_r}{k_\vartheta} \rightarrow \frac{k_r}{k_\vartheta} = s(\theta - \theta_k) - \alpha\sin\theta \equiv h(\theta),
\\
&-\frac{\nabla_\perp^2}{k_\vartheta^2} \rightarrow \frac{k_\perp^2}{k_\vartheta^2} = 1 + \frac{k_r^2}{k_\vartheta^2} = 1 + h^2(\theta) \equiv f(\theta);
\label{eq:model_salpha_grad}
\\
&\Omega_\kappa = {\bm k}_\perp\cdot(\hat{\bm b}\times{\bm \kappa}) = \frac{k_\vartheta}{R(\theta)} g(\theta), \qquad g \equiv \cos\theta + h(\theta)\sin\theta,
\label{eq:model_salpha_curv}
\end{align}
\label{eq:model_salpha}
\end{subequations}

\noindent so that $k_\perp = k_\vartheta\sqrt{f}$ and the local magnetic shear is $\partial_\theta h = s - \alpha\cos(\theta)$. Here, $\hat{\bm b} = {\bm B}/B$ and ${\bm \kappa} = \hat{\bm b}\cdot{\bm\nabla}\hat{\bm b}$ is the field line curvature vector. In this model, the equilibrium is completely described in terms of two parameters: the flux-surface-averaged magnetic shear $s$ and the normalized pressure gradient $\alpha$.

Since the CHT $s$-$\alpha$ model is derived for the low-$\beta$, large-aspect-ratio limit,\footnote{For a detailed derivation of the CHT $s$-$\alpha$ model see, e.g., Ref.~\cite{Hazeltine85}.} its application in a study of high-$\beta$ instabilities in tokamak plasmas with $\eps \gtrsim 0.1$ is somewhat ambiguous. To justify its use, it is argued that it does capture essential effects of toroidal geometry and high $\beta$; namely, toroidal curvature and modulation of the magnetic shear. The presence of these features physically distinguishes a toroidal plasma from a cylindrical or slab geometry. On the other hand, higher-order toroidal effects ignored by the $s$-$\alpha$ model (such as elliptic and triangular deformation of the flux surfaces) merely modify the lowest-order results; for instance, by adding further frequency gaps and corresponding Alfv\'{e}n eigenmodes. Thus, the $s$-$\alpha$ model is thought to be a convenient and powerful tool for basic studies of the qualitative features of low- and high-$\beta$ tokamak instabilities.

\subsection{Gyrokinetic equation}
\label{sec:model_vlasov}

The evolution of the particle distribution $\F_s$ is chosen to be governed by the Vlasov equation; i.e., the ensemble-averaged phase-space continuity equation in the absence of collisions. For a high-temperature tokamak plasma, where electromagnetic forces dominate, the Vlasov equation reads
\begin{equation}
\left[\frac{\partial}{\partial t} + {\bm v}\cdot{\bm \nabla} + \frac{e_s}{m_s}({\bm E} + {\bf v}\times{\bm B})\cdot\frac{\partial}{\partial {\bm v}} \right] \F_s = 0.
\end{equation}

\noindent Transformation into guiding center coordinates, linearization, and application of the approximations and orderings outlined in Section~\ref{sec:model_approx} yields the collisionless linear gyrokinetic equation (GKE) as derived by Antonsen \& Lane \cite{Antonsen80}. Here, we adopt the notation used by Chen \& Hasegawa (CH91) \cite{Chen91} and neglect anisotropic contributions due to $\partial_\mu f_{0s}$, where $\mu = v_\perp^2/(2B)$ is the magnetic moment. The fast gyromotion and the electron response to the parallel electric field $\delta E_\parallel = \partial_l(\dpsi - \dphi)$ are eliminated through the substitution
\begin{equation}
\delta f_s = \frac{e_s}{m_s}\left[ \left(\omega\partial_\enr\feqs\right)(\dphi - J_0^2\dpsi) - (\hat{\omega}_{*s}\feqs)J_0^2\dpsi\right] + J_0\delta K_s^{\rm CH91};
\label{eq:model_vlasov_gke1}
\end{equation}

\noindent where
\begin{align}
&\enr = v_\perp^2/2 + v_\parallel^2/2, \nonumber
\\
&\partial_l\dpsi = i\omega\delta A_\parallel \quad \text{(with Coulomb gauge)}, \nonumber
\\
&\hat{\omega}_{*s} = \omLs^{-1} ({\bm k}_\perp\times\hat{\bm b})\cdot{\bm \nabla}_{\rm g}, \nonumber
\end{align}

\noindent and ${\bm\nabla}_{\rm g}$ is the Laplacian in guiding center coordinates. Defining $\delta G_s \equiv \omega\delta K_s^{\rm CH91}$, $\delta u = \omega(\dphi-\dpsi)$, and applying the gyroaverage, one obtains
\begin{equation}
\left[v_\parallel \partial_l - i(\omega - \omegads) \right] \delta G_s = i \frac{e_s}{m_s} \left(\hat{Q}\feqs\right) (\delta S_{1s} - i\sgnVp \delta S_{2s});
\label{eq:model_vlasov_gke2}
\end{equation}

\noindent with the source terms
\begin{equation}
\delta S_{1s} = J_{0s} \delta u + \omegads J_{0s} \dpsi + \frac{v_\perp}{k_\perp} J_{1s} \omega \dBp, \qquad
\delta S_{2s} = -|v_\parallel|(\partial_l\lambda) J_1 \dpsi;
\end{equation}

\noindent where $J_i(\lambda_s)$ is the Bessel function of $i$-th order, $\lambda_s = k_\perp v_\perp / \omLs$, and $\sgnVp = v_\parallel/|v_\parallel|$. The magnetic drift frequency is
\begin{equation}
\omegads = \Omega_\kappa v _\parallel^2 / \omLs + \Omega_B \mu B / \omLs = \Omega_\kappa (v _\parallel^2 + \mu B) / \omLs + \Omega_p \mu B / \omLs, \nonumber
\end{equation}

\noindent with
\begin{align}
&\Omega_\kappa = {\bm k}_\perp\cdot(\hat{\bm b}\times{\bm \kappa}), \qquad \Omega_B = {\bm k}_\perp\cdot(\hat{\bm b}\times\nabla\ln B) \nonumber
\\
&\Omega_p = -(\mu_0/B^2) {\bm k}_\perp \cdot (\hat{\bm b}\times\nabla P). \nonumber
\end{align}

\noindent The phase-space-gradient operator $\hat{Q} = \omega\partial_\enr + \hat{\omega}_{*s}$ may be written as
\begin{equation}
\hat{Q} \rightarrow Q = (\omega_{*s}^T - \omega \Theta_s)/T_s, \nonumber
\end{equation}

\noindent where
\begin{equation}
\omega_{*s}^T = -\omega_{*s} L_{ns} \partial_r\ln\feqs\;
\stackrel{\FM}{\rightarrow}\; \omega_{*s}\left[1 + \eta_s\left(\frac{\enr}{\Ts} - \frac{3}{2}\right)\right], \quad
\Theta_s = -\T_s\partial_\enr\ln\feqs\;
\stackrel{\FM}{\rightarrow}\; 1.
\end{equation}

\noindent Here,
\begin{equation}
\omega_{*s} = \frac{k_\vartheta\Ts}{\omLsO L_{ns}}, \qquad
L_{ns}^{-1} = -\frac{\nsO'}{\nsO}, \qquad \eta_s = \frac{\Ts'/\Ts}{\nsO'/\nsO}, \nonumber
\end{equation}

\noindent and the prime denotes a radial derivative. Note that we have used the sign convention $\hat{\bm \zeta} \cdot \hat{\bm b} < 0$, so that
\begin{equation}
({\bm k}_\perp\times\hat{\bm b})\cdot{\bm\nabla}_{\rm g} \approx -k_\vartheta\partial_r. \nonumber
\end{equation}

\noindent Finally, application of the ballooning transform ($q R_0 \partial_l = \partial_\vartheta \rightarrow \partial_\theta$) and the CHT $s$-$\alpha$ model yields
\begin{equation}
\left[\frac{v_\parallel}{q R_0} \partial_\theta - i(\omega - \omegads) \right] \delta G_s = i \frac{e_s\feqs}{m_s\Ts} \left(\omega_{*s}^T - \omega \Theta_s\right) (\delta S_{1s} - i\sgnVp \delta S_{2s});
\label{eq:model_vlasov_gke3}
\end{equation}

\noindent where the coefficients in the magnetic drift frequency $\omegads$ are given by
\begin{equation}
\Omega_\kappa(\theta) = \frac{k_\vartheta}{R(\theta)} g(\theta), \qquad
\Omega_p = -\frac{k_\vartheta \alpha}{2 q^2 R(\theta) \hat{B}^2(\theta)}.
\end{equation}

For passing electrons, Eq.~(\ref{eq:model_vlasov_gke3}) is dominated by the $v_\parallel\partial_\theta$ term alone, which allows us to set $\delta G_{\rm e,pass} = 0$. As mentioned above, in the present work, we ignore trapped-electron effects and set $\delta G_{\rm e,trap} = 0$. Thus, Eq.~(\ref{eq:model_vlasov_gke3}) is solved only for ion species, while electrons are treated as a massless fluid.

\subsection{Electromagnetic field equations}
\label{sec:model_maxw}

The evolution of the electromagnetic fields is governed by Maxwell's equations. We adopt the field equations as derived by Zonca \& Chen \cite{Zonca06}, including terms containing $\partial_\theta\lambda_s$ and neglecting anisotropic contributions due to $\partial_\mu f_{0s}$. After application of the ballooning transform, the equations read
\begin{subequations}
\begin{align}
0 =& \frac{k_\vartheta^2}{(q R_0)^2} \frac{\partial}{\partial\theta}\left(f \frac{\partial\delta\psi}{\partial\theta}\right)
- \mu_0 \sum\limits_s \left<\frac{e^2}{m}(1 - J_0^2) Q f_0 \right>_s \omega\delta\phi
- \Omega_p\left[(\Omega_p + 2\Omega_\kappa)\delta\psi + \omega\delta B_\parallel\right] \nonumber
\\
&- \mu_0\sum\limits_s \left<\frac{e^2}{m}\omegad(1 - J_0^2) Q f_0 \right>_s \delta\psi
- \frac{\mu_0}{B}\sum\limits_s \left<e \mu B \left( 1 - \frac{2J_1}{\lambda} J_0\right) Q f_0 \right>_s \omega\delta B_\parallel \nonumber
\\
&- \mu_0\sum\limits_s \left<e\omegad J_0\delta G \right>_s
+ i\mu_0\sum\limits_s \left<e\frac{v_\parallel}{q R_0}(\partial_\theta\lambda) J_1\delta G \right>_s,
\label{eq:model_maxw_vort1}
\\
0 =& \sum\limits_s \left<e J_0\delta G \right>_s
+ \sum\limits_s \left<\frac{e^2}{m} \partial_{\mathcal E} f_0\right>_s \omega(\delta\phi - \delta\psi) 
+ \sum\limits_s \left<\frac{e^2}{m} (1 - J_0^2) Q f_0\right>_s \delta\psi,
\label{eq:model_maxw_qn1}
\\
0 =& \omega\delta B_\parallel + \Omega_p\delta\psi
+ \sum\limits_s \left<e \mu B\left(1 - \frac{2J_1}{\lambda}J_0\right) Q f_0\right>_s \delta\psi
+ \frac{\mu_0}{B} \sum\limits_s \left<m \mu B \frac{2J_1}{\lambda}\delta G \right>_s;
\label{eq:model_maxw_bp1}
\end{align}
\label{eq:model_maxw1}
\end{subequations}

\noindent where
\begin{equation}
Q = (\omega_{*s}^T - \omega\Theta_s)/\Ts, \qquad
\left<...\right>_s = \int{\rm d}^3 v. \nonumber
\end{equation}

\noindent Equation~(\ref{eq:model_maxw_vort1}) is the so-called vorticity equation and is obtained by substituting the parallel Amp\`{e}re's law into the continuity equation, which, in turn, is obtained by taking the zeroth-order velocity moment of the gyrokinetic equation (\ref{eq:model_vlasov_gke3}). Equation~(\ref{eq:model_maxw_qn1}) is the quasi-neutrality condition, which originally read
\begin{equation}
\sum_s\left<e_s\feqs\right> = 0, \nonumber
\end{equation}

\noindent and Eq.~(\ref{eq:model_maxw_bp1}) is the perpendicular Amp\`{e}re's law.

\subsection{Definitions and normalization}
\label{sec:model_norm}

The following dimensionless parameters are used in the present work:
\begin{align}
&\eps = \frac{r}{R_0}, \qquad
\eps_{ns} = \frac{L_{ns}}{R_0} = \eps \frac{{\rm d}\ln r}{{\rm d}\ln \nsO}, \qquad
\eps_{ps} = \eps \frac{{\rm d}\ln r}{{\rm d}\ln P_s}, \nonumber
\\
&\eta_s = \frac{{\rm d}\ln\Ts}{{\rm d}\ln\nsO}, \qquad
\tau_{ss'}^T = \frac{m_s\T_s}{m_{s'}\T_{s'}}, \qquad
\tau_{ss'}^n = \frac{Z_s\nsO}{Z_{s'} n_{0s'}}, \qquad
Z_s = -\frac{e_s}{e_{\rm e}},
\label{eq:def}
\\
&M_s = \frac{m_s}{\mi}, \qquad
\ks = \frac{k_\vartheta \Ts^{1/2}}{\omLs}, \qquad
b_s = f\ks^2, \qquad
\lambda_s = \sqrt{f}\frac{k_\vartheta v_\perp}{\omLs}. \nonumber
\end{align}

\noindent Since we are dealing with a quasi-neutral Deuterium plasma ($Z_{\rm i}=1$) containing only a sparse population of energetic particles, we have $\eps_{n{\rm i}} = \eps_{n{\rm e}}$. Note that $\eps_{ns} = \eps_{ps}(1 + \eta_s)$.

The equations are normalized in two steps. In the first step, we let
\begin{align}
&\hat{v} = \frac{v}{\vAO}, \qquad
\hat{\omega} = \frac{\omega}{\omAO}, \qquad
\hat{\Omega}_{\kappa s} = \frac{\vAO^2 \Omega_{\kappa}}{\omAO\omLs} = \frac{q\ksO g(\theta)}{\T_s^{1/2}}, \nonumber
\\
&\hat{\Omega}_{p s} = \frac{\vAO^2 \Omega_p}{\omAO\omLs} = -\frac{\alpha\ksO}{2q\T_s^{1/2}\hat{B}^2(\theta)}, \qquad
\left\{\begin{array}{c}\delta \hat{G}_s \\ \hat{F}_{{\rm M}s}\end{array}\right\} = \frac{\vAO^3}{\nsO} \left\{\begin{array}{c}\delta G_s \\ \FMs\end{array}\right\},
\label{eq:norm1}
\\
&\dPsie = \frac{e_{\rm e} \delta\psi}{\me\Te}, \qquad
\dUe = \frac{e_{\rm e}}{\me\Te} \frac{\delta u}{\omAO}, \qquad
\dCe = \frac{e_{\rm e}}{\me\Te} \frac{\vAO^2 \omega\dBp}{\omAO\omLi}; \nonumber
\end{align}

\noindent where
\begin{equation}
\vAO = \vA / \hat{B}, \qquad
\omAO = \vAO / (q R_0), \qquad
\hat{B} = B/B_0 = R_0/R. \nonumber
\end{equation}

\noindent Note that $2\hat{\T}_{\rm i} = \beta_{\rm i} = 2\mu_0 P_{\rm i}/B_0^2$ and $\ks = \ksO/\hat{B}$. In the second step, we let
\begin{align}
&\widetilde{v}^2 = \frac{\hat{v}^2}{\hat{\T}_s}, \qquad
\widetilde{\Omega}_{\kappa s} = \hat{\T}_s\hat{\Omega}_{\kappa s}, \qquad
\widetilde{\Omega}_{ps} = \hat{\T}_s\hat{\Omega}_{ps},
\label{eq:norm2}
\\
&\delta\widetilde{C}_{\rm e} = \hat{\Ti}\dCe, \qquad
\left\{\begin{array}{c}\delta\widetilde{G}_s \\ \widetilde{F}_{{\rm M}s} \end{array}\right\} = \hat{\Ts}^{3/2} \left\{\begin{array}{c}\delta\hat{G}_s \\ \hat{F}_{{\rm M}s} \end{array}\right\}; \nonumber
\end{align}

\noindent so that the normalized drift frequencies become
\begin{align}
&\widetilde{\omega}_{*s} = \frac{q\ksO\hat{\T}_s^{1/2}}{\eps_{ns}}, \qquad
\widetilde{\omega}_{{\rm d}s} = \widetilde{\Omega}_{\kappa s} \left(v_\parallel^2 + \mu B\right) + \widetilde{\Omega}_{ps} \mu B,
\\
&\widetilde{\Omega}_{\kappa s} = q\ksO\hat{\T}_s^{1/2}g(\theta), \qquad
\widetilde{\Omega}_{ps} = -\frac{\alpha \ksO\hat{\T}_s^{1/2}}{2q\hat{B}^2(\theta)}. \nonumber
\end{align}

\noindent Furthermore, we have
\begin{equation}
\lambda_s = \sqrt{b_s}\widetilde{v}_\perp, \qquad
\Theta_s = -\partial_{\widetilde{\enr}}\ln f_{0s}, \qquad
\widetilde{\omega}_{*s}^T(\FM) = \widetilde{\omega}_{*s}[1 + \eta_s(\widetilde{\enr}-3/2)]. \nonumber
\end{equation}

\noindent In the following, for simplicity, we will omit the tildes and hats indicating normalized quantities, except for $\hat{B}$ and $\ks$.

The second normalization is motivated as follows. First, the values of input parameters determining the velocity space domain size to be sampled numerically become independent of the species' temperature and may, thus, be kept constant once a set of suitable values has been determined. Second, we prevent $\Omega_{\kappa {\rm i}}$ and $\Omega_{p {\rm i}}$ from diverging, and $F_{\rm Mi}$ from collapsing into a Dirac $\delta$ distribution in the limit $\Ti \rightarrow 0$.

Note that, in \textsc{awecs}, $\Ti$ must remain finite; the case $\Ti = 0$ can only be realized if $\Ti^{1/2}$ is included in the time normalization, which is only done in an electrostatic version of the code (see Section~\ref{sec:bench_esitg}). The wave number $\ks$ is taken to be parametrically independent of $\T_s$, which implies that $k_\vartheta \propto \T_s^{-1/2}$.

All moments in Eq.~(\ref{eq:model_maxw1}) involving the Maxwellian distribution, $\left<...\FMs\right>$, can be evaluated analytically, which gives rise to the following quantities:
\begin{align}
&\Gamma_{0s} = e^{-b_s} I_0(b_s), \qquad
\Delta_{0s} = 1 - \frac{I_{1s}}{I_{0s}}, \qquad
\Delta_{1s} = 1 + \frac{b_s}{2}\left(\frac{I_{1s}}{I_{0s}} - 1\right),
\label{eq:def_moments}
\\
&\Delta_{2s} = \frac{7}{4} + \frac{b_s}{2}\left(\frac{5}{2}\frac{I_{1s}}{I_{0s}} - 3\right) - \frac{b_s^2}{2}\left(\frac{I_{1s}}{I_{0s}} - 1\right), \qquad
\Delta_{3s} = 1 + \frac{b_s}{2} \left(3\frac{I_{1s}}{I_{0s}} - 4\right) - b_s^2 \left( \frac{I_{1s}}{I_{0s}} - 1 \right), \nonumber
\\
&\Upsilon_{0s} = \Delta_{0s} + \eta_s \left(1 - 2 b_s\Delta_{0s}\right) = \Delta_{0s} + 2(\Upsilon_{1s} - 1) + \eta_s, \qquad
\Upsilon_{1s} = 1 + \eta_s b_s \left(\frac{I_{1s}}{I_{0s}} - 1\right), \nonumber
\\
&\Upsilon_{2\kappa s} = \Delta_{1s} + \eta_s \Delta_{3s}, \qquad
\Upsilon_{2ps} = \left(\Delta_{1s} - \frac{1}{2}\right) + \eta_s \left[ \Delta_{3s} - \left(\Delta_{1s} - \frac{1}{2}\right) \right]; \nonumber
\end{align}

\noindent with $I_i$ being the modified Bessel function $i$-th order. These quantities are used in the next section.

\subsection{Final form of the equations}
\label{sec:model_final}

In this section, we write down the normalized equations in a form suitable for numerical solution as an initial value problem. In the following, we denote $\partial_\theta f$ by $f'$ and the Laplace transform in time is undone. For energetic ions, the normalized velocity space integrals are given in both the general form valid for any equilibrium distribution, $F_{0s} = \feqs / \nsO$, and the form for a Mawellian ($F_{0s} = \FMs$).
\subsubsection{Marker motion and gyrokinetic equation}
\label{sec:model_final_gke}

The evolution of the position $\theta_j$ of a particle labeled $j$ is governed by the equation
\begin{equation}
{\rm d}_t \theta_j = \T_s^{1/2} v_{\parallel j}(\theta_j).
\end{equation}

\noindent In order to eliminate $\omega \leftrightarrow i\partial_t$ on the right-hand side of the GKE, the slow response of the ions is split as follows:\footnote{Before normalization, this transformation reads $\delta G_s = \delta g_s - \frac{e_s}{m_s}\left(\partial_\enr\feqs\right) \left(\delta S_{1s} - i \sgnVp\delta S_{2s}\right)$.}
\begin{equation}
\delta G_s = \delta g_s + \Theta_s F_{0s} \left(\delta S_{1s} - i \sgnVp\delta S_{2s}\right),
\label{eq:model_final_gke_dGdg}
\end{equation}

\noindent which yields (for $s \neq {\rm e}$)
\begin{equation}
\L_0 \delta g_s = -i\omegads\delta g_s + i F_{0s} \left[ (\omega_{*s}^T - \Theta_s\omegads) (\delta S_{1s} - i\hat{\sigma} \delta S_{2s}) + \Theta_s\delta\Lambda_{1s} + \Theta_s\delta\Lambda_{2s} \right].
\label{eq:model_final_gke_dg}
\end{equation}

\noindent Here, $\L_0 = \partial_t + \T_s^{1/2} v_\parallel\partial_\theta$ is the propagator and the source terms are
\begin{equation}
\delta S_{1s} =
J_{0s} \delta U_s + \omegads J_{0s} \delta \Psi_s + \frac{\tausi^T}{Z_s} \frac{v_\perp^2}{\lambda_s} J_{1s} \delta C_s, \qquad
\delta S_{2s} = -\T_s^{1/2} |v_\parallel|\lambda_s' J_{1s} \delta \Psi_s.
\end{equation}

\noindent The derivatives $\delta\Lambda_{1s} = i v_\parallel\partial_\theta\delta S_{1s}$ and $\delta\Lambda_{2s} = |v_\parallel| \partial_\theta\delta S_{2s}$ are given by
\begin{subequations}
\begin{align}
\delta\Lambda_{1s} =&
i \T_s^{1/2} v_\parallel \left\{ J_0 \delta U' + \omegad J_0 \delta\Psi' + \omegad' J_0 \delta\Psi + \frac{\tausi^T}{Z} \frac{v_\perp^2}{\lambda} J_1 \delta C' + \frac{\hat{B}'}{\hat{B}} \frac{\tausi^T}{Z} \frac{v_\perp^2}{\lambda} J_1 \delta C \right\}_s
\\
&
+ i \T_s^{1/2} v_\parallel \lambda'_s \left\{ -J_1 \delta U - \omegad J_1 \delta\Psi + \frac{\tausi^T}{Z} \frac{v_\perp^2}{\lambda} \left(J_0 - \frac{2 J_1}{\lambda}\right) \delta C \right\}_s, \nonumber
\\
\delta\Lambda_{2s} =& T_s \mu B \frac{B'}{B} \lambda'_s J_{1s} \delta\Psi_s
- \T_s v_\parallel^2 \left\{ \lambda' J_1 \delta\Psi' + \lambda \left[\frac{\lambda''}{\lambda} - \left(\frac{\lambda'}{\lambda}\right)^2 \right] J_1\delta\Psi + (\lambda')^2 J_0 \delta\Psi \right\}_s;
\end{align}
\end{subequations}

\noindent where $\delta S_s = -Z_s\taues^T \delta S_{\rm e}$ (similarly for $\delta \Psi_s$, $\delta U_s$, and $\delta C_s$), $J_i = J_i(\lambda)$, and $\lambda_s = \sqrt{f}\ks v_\perp$. The derivatives of $\lambda_s$ and $\omegads$ with respect to $\theta$ are
\begin{align}
&\lambda_s'/\lambda = h(s - \alpha\cos\theta)/f - \hat{B}'/(2\hat{B}), \nonumber
\\
&\lambda_s''/\lambda_s - (\lambda'/\lambda)^2 = [h \alpha \sin\theta + (2-f)(h')^2] / f^2 + \hat{B}''/(2\hat{B}) - (\hat{B}'/\hat{B})^2/2, \nonumber
\\
&\omegads' = (\Omega_{\kappa s}/g)[h\cos\theta + (h'-1)\sin\theta](v_\parallel^2 + \mu B) - (\Omega_{\kappa s} + \Omega_{ps}) \mu B (\hat{B}'/\hat{B}).
\end{align}

\subsubsection{Vorticity equation}
\label{sec:model_final_vort}

The vorticity equation is a second-order differential equation in time $t$. Collecting all time derivatives in an auxiliary field $\dEe$, we obtain two first-order equations: the continuity equation\footnote{Note that we have combined the drift-kinetic and FLR components of the ballooning terms ($\propto \dPsie,\dCe$) which are separate in Eq.~(\protect\ref{eq:model_maxw_vort1}).}
\begin{align}
\partial_t \dEe
=& \kiO^2 \left[f \dPsie'' + 2hh' \dPsie'\right]
- \left[ \omega_{*{\rm i}} \left(1 - \Gamma_{0{\rm i}}\Upsilon_{1{\rm i}}\right) + H_{EU} \right] \dUe
\label{eq:model_final_de}
\\
& + \omega_{*{\rm i}} \left[ 2\tauei^T(1 + \eta_{\rm e}) \Omega_{\kappa{\rm i}} + \tauei^T(1 + \eta_{\rm e}) \Omega_{p{\rm i}} + 2\Omega_{\kappa{\rm i}} \Gamma_{0{\rm i}} \Upsilon_{2\kappa{\rm i}} + 2\Omega_{p{\rm i}} \Gamma_{0{\rm i}} \Upsilon_{2p{\rm i}}\right] \dPsie \nonumber
\\
& + H_{E\Psi} \dPsie
+ \omega_{*{\rm i}}\left[\tauei^T(1 + \eta_{\rm e}) + \Gamma_{0{\rm i}}\Upsilon_{0{\rm i}}\right] \dCe
+ H_{EC} \dCe \nonumber
\\
&+ \frac{1}{\tauei^T} \left<\omegad J_0 \delta G\right>_{\rm i}
+ \frac{\tauEi^n}{\tauei^T} \left<\omegad J_0 \delta G\right>_{\rm E}
- i\frac{\Ti^{1/2}}{\tauei^T} \left<v_\parallel \lambda' J_1 \delta G\right>_{\rm i}
- i\frac{\TE^{1/2}\tauEi^n}{\tauei^T} \left<v_\parallel \lambda' J_1 \delta G\right>_{\rm E}, \nonumber
\end{align}

\noindent and the parallel Amp\`{e}re's law (which now defines $\dEe$)
\begin{align}
A_\omega \partial_t \dPsie =&
i \left[ 2(\Omega_{\kappa{\rm i}} + \Omega_{p{\rm i}})(1 - \Gamma_{0{\rm i}}\Delta_{1{\rm i}}) - \Omega_{p{\rm i}}(1 - \Gamma_{0{\rm i}}) - \omega_{*{\rm i}} (1 - \Gamma_{0{\rm i}}\Upsilon_{1{\rm i}}) \right] \dPsie
\label{eq:model_final_da}
\\
&+ i H_{\Psi\Psi} \dPsie
+ \dEe
+ i A_\omega \dUe 
+ i (1 - \Gamma_{0{\rm i}}\Delta_{0{\rm i}}) \dCe
+ i H_{\Psi C} \dCe; \nonumber
\end{align}

\noindent where $A_\omega = (1 - \Gamma_{0{\rm i}}) + H_\omega$, and the energetic particle terms are
\begin{align}
H_{EU} =& Z_{\rm E} \tauEi^n \tauiE^T \left<(1 - J_0^2) \omega_*^T F_0 \right>_{\rm E}
\stackrel{\FM}{\rightarrow} Z_{\rm E} \tauEi^n \tauiE^T \omega_{*{\rm E}} \left(1 - \Gamma_{0{\rm E}}\Upsilon_{1{\rm E}}\right),
\\
H_{E\Psi} =& Z_{\rm E} \tauEi^n\tauiE^T \left<J_0^2 \omegad \omega_*^T F_0\right>_{\rm E}
\stackrel{\FM}{\rightarrow} 2 \tauEi^n \omega_{*{\rm E}} \left[ \Omega_{\kappa{\rm i}} \Gamma_{0{\rm E}} \Upsilon_{2\kappa{\rm E}} + \Omega_{p{\rm i}} \Gamma_{0{\rm i}} \Upsilon_{2p{\rm E}}\right], \nonumber
\\
H_{EC} =& \tauEi^n\left<\mu B \frac{2J_1 J_0}{\lambda} \omega_*^T F_0 \right>_{\rm E}
\stackrel{\FM}{\rightarrow} \tauEi^n\omega_{*{\rm E}} \Gamma_{0{\rm E}}\Upsilon_{0{\rm E}}, \nonumber
\\
H_{\Psi\Psi} =& Z_{\rm E} \tauEi^n\tauiE^T \left<(1 - J_0^2) \omegad \Theta_s F_0\right>_{\rm E}
- H_{EU}, \nonumber
\\
\stackrel{\FM}{\rightarrow}& 2\tauEi^n(\Omega_{\kappa{\rm i}} + \Omega_{p{\rm i}})(1 - \Gamma_{0{\rm E}}\Delta_{1{\rm E}}) - \tauEi^n\Omega_{p{\rm i}}(1 - \Gamma_{0{\rm E}}) - H_{EU}(F_0=\FM), \nonumber
\\
H_{\Psi C} =& \tauEi^n\left<\mu B\left(1 - \frac{2 J_1 J_0}{\lambda}\right) \Theta_s F_0\right>_{\rm E}
\stackrel{\FM}{\rightarrow} \tauEi^n(1 - \Gamma_{0{\rm E}}\Delta_{0{\rm E}}), \nonumber
\\
H_\omega =& Z_{\rm E} \tauEi^n \tauiE^T \left<(1 - J_0^2) \Theta_s F_0\right>_{\rm E}
\stackrel{\FM}{\rightarrow} Z_{\rm E}\tauEi^n\tauiE^T(1 - \Gamma_{0{\rm E}}). \nonumber
\end{align}

\noindent The moments of the transformation $\delta G \rightarrow \delta g$ are:
\begin{align}
\frac{\left<\omegad J_0 (\delta G-\delta g)\right>_s}{\tauei^T} =&
- Z_s\tauis^T \left[ \left<\omegad J_0^2 \Theta F_0\right>_s\dUe
+ \left<\omegad^2 J_0^2 \Theta F_0\right>_s\dPsie \right]
- \left<\omegad \mu B \frac{2 J_1 J_0}{\lambda} \Theta F_0\right>_s\dCe \nonumber
\\
\stackrel{\FM}{\rightarrow}&
- \Gamma_{0s} \left[ 2 \left(\Omega_{\kappa{\rm i}} + \Omega_{p{\rm i}}\right)\Delta_{1s} - \Omega_{p{\rm i}}\right] \dUe \nonumber
\\
&- 2 \frac{\tausi^T}{Z_s} \Gamma_{0s} \left[ (\Omega_{\kappa{\rm i}} + \Omega_{p{\rm i}})^2 \Delta_{3s} + 2\Omega_{\kappa{\rm i}}(\Omega_{\kappa{\rm i}} + \Omega_{p{\rm i}}) \left(\Delta_{1s} - \frac{1}{2}\right) + \frac{3}{2} \Omega_{\kappa{\rm i}}^2 \right] \dPsie \nonumber
\\
&- \frac{\tausi^T}{Z_s} \Gamma_{0s} \left[ \Omega_{\kappa{\rm i}} (1 + 2\Delta_{0s} - 2b_{\rm i}\Delta_{0s}) + \Omega_{p{\rm i}} (1 +  \Delta_{0s} - 2b_{\rm i}\Delta_{0s}) \right] \dCe,
\label{eq:model_final_dh1}
\end{align}

\vspace*{-0.4cm}
\begin{align}
- i\frac{\sqrt{\T_s}}{\tauei^T} \left<v_\parallel \lambda' J_1 (\delta G-\delta g)\right>_s =&
- \frac{Z_s}{M_s} \Ti \left(\frac{\lambda'_s}{\lambda_s}\right)^2 \left<v_\parallel^2 \lambda^2 J_1^2 \Theta F_0\right>_s \dPsie \nonumber
\\
\stackrel{\FM}{\rightarrow}&
- 2 \frac{Z_s}{M_s} \Ti \left(\frac{\lambda'_s}{\lambda_s}\right)^2 b_s^2 \Gamma_{0s} \Delta_{0s} \dPsie.
\label{eq:model_final_dh2}
\end{align}

\noindent In the low-$\beta$ limit we may set
\begin{equation}
\dBp = \Omega_p = 0. \nonumber
\end{equation}

\noindent By neglecting the energetic ion terms we then recover the model used by Zhao \& Chen, 2002 \cite{Zhao02}. In this case, the term on the second line of Eq.~(\ref{eq:model_final_dh1}) reduces to
\begin{equation}
\Omega_\kappa^2[\Delta_3 + 2(\Delta_1-1/2) + 3/2] = 2\Omega_\kappa^2\Delta_2. \nonumber
\end{equation}

\noindent Note that our calculation yields a different $\Delta_2$ than that given in Ref.~\cite{Zhao02}: $\Delta_2^{\rm Zhao} = 5/2 + ...$, whereas here $\Delta_2 = 7/4 + ...$ [cf.~Eq.~(\ref{eq:def_moments})]. Our $\Delta_2$ gives slightly larger growth rates, as will be shown in the benchmark in Fig.~\ref{fig:bench_saw_zhao} below.

\subsubsection{Quasi-neutrality condition}
\label{sec:model_final_qn}

After eliminating $\partial_t\dPsie$ with the help of Eq.~(\ref{eq:model_final_da}), the quasi-neutrality condition becomes an algebraic equation, which may be written as
\begin{align}
\left[\frac{1}{\tauei^T} + \Gamma_{0{\rm i}} -(H_\omega - Z_{\rm E}\tauEi^n\tauiE^T)\right] \dUe =& -i\dEe
+ \left[ 2(\Omega_{\kappa{\rm i}} + \Omega_{p{\rm i}}) (1 - \Gamma_{0{\rm i}} \Delta_{1{\rm i}}) - \Omega_{p{\rm i}} (1 - \Gamma_{0{\rm i}})\right] \dPsie \nonumber
\\
& + H_{U\Psi}\dPsie
+ (1 - \Gamma_{0{\rm i}} \Delta_{0{\rm i}}) \dCe
+ H_{UC}\dCe \nonumber
\\
& - \frac{1}{\tauei^T} \left<J_0 \delta G\right>_{\rm i}
- \frac{\tauEi^n}{\tauei^T} \left<J_0 \delta G\right>_{\rm E}.
\label{eq:model_final_du}
\end{align}

\noindent The energetic particle terms are
\begin{align}
H_{U\Psi} =& Z_{\rm E} \tauEi^n \tauiE^T \left<(1 - J_0^2)\omegad \Theta F_0 \right>_{\rm E}
\;\stackrel{\FM}{\rightarrow}\; \tauEi^n \left[ 2(\Omega_{\kappa{\rm i}} + \Omega_{p{\rm i}}) (1 - \Gamma_{0{\rm E}} \Delta_{1{\rm E}}) - \Omega_{p{\rm i}} (1 - \Gamma_{0{\rm E}})\right], \nonumber
\\
H_{UC} =& \tauEi^n \left<\left(1 - \frac{2 J_1 J_0}{\lambda}\right) \mu B \Theta F_0 \right>_{\rm E}
\;\stackrel{\FM}{\rightarrow}\; \tauEi^n (1 - \Gamma_{0{\rm E}} \Delta_{0{\rm E}}).
\end{align}

\noindent The moments of the transformation $\delta G \rightarrow \delta g$ are:
\begin{align}
-\frac{\left< J_0 (\delta G-\delta g)\right>_s}{\tauei^T} =&
\frac{Z_s \taues^T}{\tauei^T} \left[ \left<J_0^2 \Theta F_0\right>_s \dUe
+ \left<\omegad J_0^2 \Theta F_0\right>_s \dPsie \right]
+ \frac{1}{\tauei^T} \left<\mu B \frac{2 J_1 J_0}{\lambda} \Theta F_0\right>_s \dCe \nonumber
\\
\stackrel{\FM}{\rightarrow}&\;
Z_s \tauis^T \Gamma_{0s} \dUe
+ \Gamma_{0s} \left[2\Omega_{\kappa{\rm i}} \Delta_{1s} + \Omega_{p{\rm i}}\left( 2\Delta_{1s} - 1 \right)\right] \dPsie
+ \Gamma_{0s} \Delta_{0s} \dCe.
\label{eq:model_final_dh3}
\end{align}

\noindent Given the orderings described in Section~\ref{sec:model_approx}; in particular, $\tauiE^T \sim \tauEi^n \sim \O(\delta^2...\delta)$, we expect the contribution of energetic ions to the quasi-neutrality condition to be small.

\subsubsection{Magnetic compression}
\label{sec:model_final_bp}

After eliminating $\partial_t\dPsie$ with the help of Eq.~(\ref{eq:model_final_da}), the perpendicular Amp\`{e}re's law becomes an algebraic equation, which may be written as
\begin{align}
\left(\hat{B}^2 + \Ti A_\Sigma^2 A_\omega\right) \dCe =& -i \Ti A_\Sigma \dEe
+ \Ti A_\Sigma A_\omega \dUe
+ \Ti A_\Sigma H_{\Psi\Psi} \dPsie \nonumber
\\
&+ \Ti A_\Sigma \left[ 2(\Omega_{\kappa{\rm i}} + \Omega_{p{\rm i}})(1 - \Gamma_{0{\rm i}}\Delta_{1{\rm i}}) - \Omega_{p{\rm i}}(1 - \Gamma_{0{\rm i}}) - \omega_{*{\rm i}}(1 - \Gamma_{0{\rm i}}\Upsilon_{1{\rm i}}) \right] \dPsie \nonumber
\\
&+ \Ti \omega_{*{\rm i}} \tauei^T (1 + \eta_{\rm e}) \dPsie
+ \Ti \omega_{*{\rm i}} \Gamma_{0{\rm i}} \left[ \Delta_{0{\rm i}} + \eta_{\rm i} \left(1 - 2 b_{\rm i} \Delta_{0{\rm i}}\right) \right] \dPsie
+ H_{C\Psi}\dPsie \nonumber
\\
&+ \frac{\Ti}{\sqrt{b_{\rm i}} \tauei^T} \left< v_\perp J_1 \delta G \right>_{\rm i}
+ \frac{\Ti\tauEi^n}{\sqrt{b_{\rm E}} Z_{\rm E} \tauei^T \tauiE^T} \left< v_\perp J_1 \delta G \right>_{\rm E};
\label{eq:model_final_dc}
\end{align}

\noindent where
\begin{equation}
A_\Sigma = -\left[(1 - \Gamma_{0{\rm i}} \Delta_{0{\rm i}}) + H_{UC}\right] / A_\omega. \nonumber
\end{equation}

\noindent The energetic particle term $H_{C\Psi}$ is
\begin{align}
H_{C\Psi} =& - \Ti\tauEi^n \left< \frac{2 J_1 J_0}{\lambda} \mu B \omega_*^T F_0 \right>_{\rm E} \stackrel{\FM}{\rightarrow}
\Ti \tauEi^n \omega_{*{\rm E}} \Gamma_{0{\rm E}} \left[ \Delta_{0{\rm E}} + \eta_{\rm E} \left(1 - 2 b_{\rm E} \Delta_{0{\rm E}} \right)\right].
\end{align}

\noindent The moments of the transformation $\delta G \rightarrow \delta g$ are:
\begin{align}
\Ti\frac{\left< v_\perp J_1 (\delta G-\delta g)\right>_s}{Z_s \tauei^T \tauis^T\sqrt{b_s}}
=&
- \frac{\Ti}{\sqrt{b_s}} \left[ \left<v_\perp J_0 J_1 \Theta F_0\right>_s \dUe
+ \left<\omegad v_\perp J_0 J_1 \Theta F_0\right>_s \dPsie \right] \nonumber
\\
& - \frac{\Ti}{\sqrt{b_s}} \frac{\tausi^T}{Z_s} \left<v_\perp \mu B \frac{2 J_1^2}{\lambda} \Theta F_0\right>_s \dCe \nonumber
\\
\stackrel{\FM}{\rightarrow}&
- \frac{\Ti \tausi^T}{Z_s} \Gamma_{0s} \left[ \Omega_{\kappa{\rm i}} (1 + 2\Delta_{0s} - 2b_s\Delta_{0s}) + \Omega_{p{\rm i}} (1 + \Delta_{0s} - 2b_s\Delta_{0s}) \right] \dPsie \nonumber
\\
&- \frac{2 \Ti \tausi^T}{Z_s} \Gamma_{0s} \Delta_{0s} \dCe
- \Ti\Gamma_{0s} \Delta_{0s} \dUe.
\label{eq:model_final_dh4}
\end{align}

\noindent The final form used in \textsc{awecs} is obtained by substituting Eq.~(\ref{eq:model_final_dc}) into Eq.~(\ref{eq:model_final_du}) to eliminate $\dCe$.

\section{Numerical methods}
\label{sec:num}

Equations (\ref{eq:model_final_gke_dg}), (\ref{eq:model_final_de}), (\ref{eq:model_final_da}), (\ref{eq:model_final_du}) and (\ref{eq:model_final_dc}) are solved as an initial value problem with the particle-in-cell (PIC) code \textsc{awecs} using a Runge-Kutta scheme. A finite number of markers is employed to sample the phase space. A modified $\delta f$ method appropriate for particle-conserving compressible dynamics is adopted, with marker weights chosen such as to allow a uniform distribution in energy. In this section, these methods are described in detail and an outline of the computational cycle is given. In the following, grid points and markers are labeled by the indices $i \in [1,N_{\rm g}]$ and $j \in [1, N_{\rm m}]$, respectively. A ``cell'' is the space between two grid points and its size is $\Delta\theta = \theta_{i+1} - \theta_i$.

\subsection{PIC method}
\label{sec:num_pic}

While marker positions $\theta_j$ vary continuously, fields are sampled at discrete grid points $\theta_i$. In order to solve the field equation, the contribution of each marker to the particle density at each grid point must be determined. Conversely, the marker motion is subject to electromagnetic forces known only on the discrete grid. The PIC method employed here is a 1st-order scheme that provides smooth mapping between markers and the grid. In this method, each marker $j$ is replaced by a top-hat function $\Pi$ of width $\Delta\theta$ centered at $\theta_j$ ,where the definition of the top-hat function is $\Pi(x) = 1$ for $|x| < 1$, and zero elsewhere. The sum of these finite-sized markers integrated over the interval $\theta \in [\theta_i-\Delta\theta/2, \theta_i+\Delta\theta/2]$ and divided by $\Delta\theta$ yields the number of markers $N_i$ contributing to grid point $\theta_i$:
\begin{equation}
\frac{1}{\Delta\theta} \sum\limits_{j=1}^{N_{\rm m}} \text{  } \int\limits_{\theta_i - \Delta\theta/2}^{\theta_i + \Delta\theta/2} {\rm d}\theta \text{  } \Pi\left(\frac{\theta-\theta_j}{\Delta\theta/2}\right) = \sum_j S(\theta_i - \theta_j) \equiv N_i.
\label{eq:markers_pic_ni}
\end{equation}

\noindent This local integration gives rise to the triangular shape function\footnote{It may be more intuitive to refer to $\Pi$ as the ``shape'' of a particle. The function $S$ describes the mapping between marker position and the finite-difference grid and is sometimes called ``assignment function.'' In practice, we only deal with $S$ and it has become customary to simply call it ``shape function.''}
\begin{equation}
S(\theta - \theta_j) = \left\{\begin{array}{lcl}
1 - 2|\theta - \theta_j|/\Delta\theta & : & |\theta - \theta_j| \leq \Delta\theta/2, \\
0 & : & \text{elsewhere}.
\end{array}\right.
\end{equation}

\noindent Note that $\int{\rm d}\theta\; S = \Delta \theta$. At the boundary points of the domain in which markers are loaded, the marker weights are doubled. This effectively simulates the effect of a plasma beyond these points, which is a mirror image of the plasma inside.

\subsection{Modified ${\bm \delta f}$ method for compressible dynamics}
\label{sec:num_df}

\subsubsection{Description of the method}
\label{sec:num_df_method}

Equation~(\ref{eq:model_vlasov_gke3}) may be compactly written as
\begin{equation}
\L_0 \delta G_s + i\omegads\delta G_s = - \delta\L \feqs, \nonumber
\end{equation}

\noindent where $\L_0 = \partial_t + v_\parallel\partial_l$ is the inverse propagator along a field line. The $\omegads$ term can be eliminated by a transformation from the guiding centers to magnetic drift centers \cite{Tsai93}
\begin{align}
\delta G_s = \dGds e^{-i\delds} \quad & \text{with} \quad \delds = \int\limits_{-\infty}^t {\rm d}\tau\; \omegads(\tau) \nonumber
\\
&\Rightarrow \quad \L_0 \dGds = - \delta\L \feqs e^{i\delds}.
\label{eq:num_gke1}
\end{align}

\noindent The integrating factor $\exp(-i\delta_{{\rm d}s})$ was expected to help avoid numerical problems associated with the secular term in $\omegads$ ($\omegads \propto \theta$ for $|\theta| \gg 1$). To date, however, no significant difference was found between \textsc{awecs} runs solving Eq.~(\ref{eq:model_vlasov_gke3}) and those solving Eq.~(\ref{eq:num_gke1}).

For an isotropic Maxwellian equilibrium distribution, $f_0 = n_0(r)\FM(r,\enr)$, the particle density at high energies $\enr$ is low, so a corresponding marker distribution would introduce a large amount of discretization noise through the highest-order velocity moment, which in our case is $\left<\omegad J_0 \delta G\right>$. One way to avoid this problem is to load the markers with a probability distribution function (PDF) $P_{\sgnVp}$, defined by $P_{\sgnVp} W_0 = f_0$, and require that $\partial_\enr P_{\sgnVp} = 0$. For the perturbed particle distribution this means that $P_{\sgnVp} \delta W = \delta f$, or, equivalently, $\delta W/W_0 = \delta f/f_0$, where $W_0$ and $\delta W$ are equilibrium and perturbed weight functions.

In the conventional $\delta f$ method, the PDF $P_{\sgnVp}$ for the markers is chosen to be such that $\L_0 P_{\sgnVp} = 0$, and one solves the equation for the weight function $\delta W$ instead of $\delta f$. In the present case, that equation would read
\begin{equation}
[\L_0 - (\partial_l v_\parallel)] \delta W = -W_0 \delta\L(\ln\feqs). \nonumber
\end{equation}

\noindent In \textsc{awecs}, we adopt the alternative scheme utilized previously in Ref.~\cite{Dettrick03}, where $P_{\sgnVp}$ is required to satisfy the continuity equation for compressible dynamics,
\begin{equation}
\partial_t P_{\sgnVp} + \partial_\theta(v_\parallel P_{\sgnVp}) = 0.
\label{eq:num_pdf}
\end{equation}

\noindent Defined this way, the spatial marker distribution represents closely the physical particle distribution along a field line. Therefore, we can solve the GKE for the physical particle distribution function,
\begin{equation}
\L_0 \dGds = -\delta\L \feqs \exp(i\delds), \nonumber
\end{equation}

\noindent along unperturbed marker orbits. The difference between markers and physical particles manifests itself only in the discretized velocity space integrals, where an additional weight factor appears (see Section~\ref{sec:num_df_dv}).

\subsubsection{Marker loading}
\label{sec:num_df_load}

It is convenient to introduce the pitch angle variable $A$ defined as
\begin{equation}
A = \frac{\mu B_0}{\enr} = \frac{v_\perp^2}{v^2} \frac{B_0}{B} = \frac{\sin^2(\varphi)}{\hat{B}};
\end{equation}

\noindent where $\varphi = \varphi(B)$ is the pitch angle; i.e., the angle between the local magnetic field ${\bf B}$ and the velocity vector ${\bf v}$. Note that for particles which are trapped in a magnetic mirror, the pitch angle coordinate $A$ is related to the turning points $\pm\theta_{\rm b}$ (``bounce angles'') through $A = 1/\hat{B}(\theta_{\rm b})$ which follows from $v_\perp^2(\theta_{\rm b}) = 2\enr$. For the purpose of marker loading, we take $A$ to be an independent velocity space coordinate instead of the magnetic moment $\mu$. Recall that
\begin{equation}
\enr = v^2/2 = (v_\perp^2 + v_\parallel^2)/2, \qquad
\mu = v_\perp^2/(2B), \nonumber
\end{equation}

\noindent so the velocity space coordinates do not contain the particle mass. For the parallel velocity $v_\parallel$ of a marker with $(A,\enr)$ at a given location $\theta_j$, parametrized by $B=B(\theta_j)$, we obtain the following expression:
\begin{equation}
|v_\parallel| = \sqrt{v^2 - v_\perp^2} = \sqrt{2\enr} \sqrt{1 - A\hat{B}} = \sqrt{2\enr} u, \qquad \text{with} \quad u = \sqrt{1 - A\hat{B}}.
\end{equation}

\noindent Note that the normalized parallel velocity $u$ satisfies $\partial_\enr u = 0$.

In order for the plasma conditions to remain constant in time, the markers must be loaded in equilibrium: $\partial_t P_{\sgnVp} = 0$. Thus, Eq.~(\ref{eq:num_pdf}) implies $\partial_l(v_\parallel P_{\sgnVp}) = 0$; that is,
\begin{equation}
v_\parallel P_{\sgnVp} = C(A,\enr). \nonumber
\end{equation}

\noindent Since $\partial_\enr A = 0$, a simple PDF which satisfies the condition $v_\parallel P_{\sgnVp} = C(A,\enr)$ and allows to initialize uniformly in energy ($\partial_\enr P_{\sgnVp} = 0$) is obtained with the choice $C(A,\enr) = C_0 \sqrt{2\enr}$:
\begin{equation}
P_{\sgnVp}(A, B) = \frac{C_0 \sqrt{2\enr}}{|v_\parallel|} = \frac{C_0}{\sqrt{1 - A\hat{B}}} = \frac{C_0}{u}.
\label{eq:markers_pdf}
\end{equation}

The size of the computational domain, $\theta \in [-\theta_{\rm max}, \theta_{\rm max}]$ must be sufficiently large to avoid unphysical reflections at the boundaries. However, markers are only required in the region where the unstable modes have a significant amplitude. In \textsc{awecs}, the parameter $N_{\rm p}$ determines the number of loading periods, and the size of this domain, $2\pi N_{\rm p}$, can usually be chosen smaller than the computational domain, $2\theta_{\rm max}$ (see the convergence study in Section~\ref{sec:accurate_converge}). In the region without markers, we set $\delta G = 0$.

The markers are loaded according to the following procedure:
\begin{enumerate}
\item  \underline{Spatial loading.} Integrating the marker PDF $P_{\sgnVp}$ given by Eq.~(\ref{eq:markers_pdf}), and using $\sum_{\sgnVp}P_{\sgnVp} = 2P_{\sgnVp}$, we obtain
\begin{equation}
\sum\limits_{\sgnVp} \int\limits_{\enr_{\rm min}}^{\enr_{\rm max}} {\rm d}\enr \int\limits_{A_{\rm min}}^{A_{\rm max}} {\rm d}A \int\limits_{\theta_1}^{\theta_2} {\rm d}\theta \; P_{\sgnVp} = 2 C_0 (\enr_{\rm max}-\enr_{\rm min}) \int\limits_{\theta_1}^{\theta_2} {\rm d}\theta \left[ -\frac{2}{\hat{B}} \sqrt{1 - A\hat{B}} \right]_{A_{\rm min}}^{A_{\rm max}};
\label{eq:markers_spatial_load1}
\end{equation}

which is valid for any interval $[\theta_1, \theta_2]$. Hence, the marker density $n(\theta)$ is given by
\begin{equation}
n(\theta) \propto w(\theta) = \frac{1}{\hat{B}} \left( \sqrt{1 - A_{\rm min}\hat{B}} - \sqrt{\text{max}\{1 - A_{\rm max}\hat{B}, 0\}} \right) = \frac{u_{\rm max} - u_{\rm min}}{\hat{B}}.
\end{equation}

The weight function $w(\theta)$ is then used to map a uniform distribution of random numbers $R_j \in [0, 1]$ to a nonuniform distribution in $\theta$:
\begin{enumerate}
\item  Numerical integration of $w$ gives the cumulative distribution
\begin{equation}
\overline{W}(\theta) = \int_{\theta_1}^\theta{\rm d}\theta' w(\theta'). \nonumber
\end{equation}

\item  From this, we sample uniformly distributed random values,
\begin{equation}
R_j \overline{W}(\theta_2) = \overline{W}_j \in [0, \overline{W}(\theta_2)]. \nonumber
\end{equation}
In ballooning space, the integral limits are $\theta_1 = 0$ and $\theta_2 = \max\{\theta_{\rm b}\}-\eps_{\rm num}$ for trapped particles, while for passing particles $\theta_2 = \pi - \eps_{\rm num}$. Here, $\eps_{\rm num}$ is the smallest number that can be represented numerically in double precision.

\item  The map $\overline{W}^{-1}(\theta): \; \overline{W}_j \rightarrow \theta_j$ then yields non-uniformly distributed marker positions $\theta_j$. Since $\overline{W}$ is known only at discrete grid points, the inversion $\overline{W}^{-1}$ is done using linear interpolation.

\item  Offsets $n_{{\rm p}j}\pi$, with $n_{{\rm p}j} = 1,...,(N_{\rm p} - 1)$, are added to spread markers over all periods.
\end{enumerate}

\item  \underline{Pitch angle distribution:} The largest allowed value for $A$ depends on the location $\theta$ (parametrized by $B$). The constraint to be obeyed is most easily written for the parallel velocity, which must satisfy
\begin{equation}
v^2 - v_\perp^2 = v_\parallel^2 \geq 0. \nonumber
\end{equation}
Thus, we first determine the limits of $u \equiv |v_\parallel|/\sqrt{2\enr}$ for a chosen interval $[A_{\rm min}, A_{\rm max}]$, using the relation
\begin{equation}
u = \sqrt{1 - A\hat{B}}. \nonumber
\end{equation}
These are
\begin{equation}
u_{\rm max} = \sqrt{1 - A_{\rm min}\hat{B}}, \qquad
u_{\rm min} = \sqrt{\max\{1 - A_{\rm max}\hat{B}, \; 0\}}. \nonumber
\end{equation}
Next, we sample random values $u_j$, uniformly distributed over the interval $[u_{\rm min}, u_{\rm max}]$, and calculate the pitch angle variable $A_j = (1 - u_j^2)/\hat{B}$. Note that passing particles satisfy $A_{\rm max} = A_\pi$, where $A_\pi \equiv 1/\hat{B}(\pi) = 1- \eps$.

\item  \underline{Energy distribution:} We distribute marker energies uniformly in a chosen interval $[\enr_{\rm min}, \enr_{\rm max}]$. The appropriate limits depend on the problem at hand; in particular, the order of the highest energy moment. In \textsc{awecs}, the energy coordinate $\enr_j$ is also used to store the sign of the parallel velocity, $\sgnVp = \text{sign}(v_\parallel)$.
\end{enumerate}

\noindent After loading $N_{{\rm m}s}/4$ markers onto the positive $\theta$ and $\enr$ axes as described above, the distribution is copied to the respective negative axis. The number of markers used for species $s$ is thus given by $N_{{\rm m}s} = N_{{\rm f}s}\times 4 \times N_{\rm p}$, with input parameters $N_{{\rm f}s}$ and $N_{\rm p}$. The remaining input parameters specifying the marker distribution are
\begin{equation}
v_{\rm min}, \qquad
v_{\rm max}, \qquad
\underbrace{0 < a_{\pi{\rm min}} < a_{\pi{\rm max}} < 1,}\limits_{\text{passing}} \qquad
\underbrace{0 < \theta_{\rm b,min} < \theta_{\rm b,max} < \pi}\limits_{\text{trapped}},
\label{eq:markers_input}
\end{equation}

\noindent such that
\begin{equation}
\enr \in \frac{1}{2}\times [v_{\rm min}^2, v_{\rm max}^2], \qquad
A_{\rm pass} \in A_\pi\times [a_{\pi{\rm min}}, a_{\pi{\rm max}}], \qquad
A_{\rm trap} \in [1/\hat{B}(\theta_{\rm b,max}), 1/\hat{B}(\theta_{\rm b,min})]. \nonumber
\end{equation}

\noindent The fraction of trapped particles can only be manipulated through the inverse aspect ratio $\eps$. However, the number of markers used to sample the phase spaces of trapped and passing particles can be varied independently.

The marker loading is completed with the calculation of the normalization constant $C_0$. If we equate the integral of $P_{\sgnVp}$ in Eq.~(\ref{eq:markers_spatial_load1}) with that of the Klimontovich distribution,
\begin{equation}
P_\delta = \sum_{j=1}^{N_{\rm m}} \delta(\enr - \enr_j) \delta(A - A_j) \delta(\theta - \theta_j) \delta(\sgnVp - \sgnVp_j),
\label{eq:num_pdelta}
\end{equation}

\noindent we obtain the following equation for the constant $C_0$:
\begin{equation}
C_0 = \frac{N_{\rm m}(\theta_{\rm min}, \theta_{\rm max})}{4(\enr_{\rm max}-\enr_{\rm min}) b_1} \qquad \text{with} \qquad b_1 = \int\limits_{\theta_{\rm min}}^{\theta_{\rm max}} {\rm d}\theta \frac{u_{\rm max} - u_{\rm min}}{\hat{B}};
\end{equation}

\noindent where $N_{\rm m}(\theta_{\rm min}, \theta_{\rm max})$ is the number of markers in the interval $\theta \in [\theta_{\rm min}, \theta_{\rm max}]$. In the PIC method, the spatial Dirac deltas $\delta(\theta - \theta_j)$ in the Klimontovitch distribution (\ref{eq:num_pdelta}) are replaced by finite-sized markers $S(\theta - \theta_j)/\Delta\theta$ [cf.~Eq.~(\ref{eq:markers_pic_ni})]. The local normalization constant $C_{0i}$ at a grid point $i$ is obtained by carrying out the integral in Eq.~(\ref{eq:markers_spatial_load1}) over the spatial interval $\theta \in [\theta_i - \Delta\theta/2, \theta_i + \Delta\theta/2]$ and substituting Eq.~(\ref{eq:markers_pic_ni}) for the left-hand side:
\begin{equation}
\sum_{j=1}^{N_{\rm m}} S(\theta_i - \theta_j) = N_i
\approx 4 C_{0i} (\enr_{\rm max}-\enr_{\rm min}) \left[ \frac{u_{\rm max} - u_{\rm min}}{\hat{B}} \right]_i \Delta\theta;
\label{eq:markers_spacial_load2}
\end{equation}

\noindent where $\hat{B}$ was taken to be constant within the small integration interval. As in Eq.~(\ref{eq:markers_pic_ni}), $N_i$ denotes the effective number of markers at the grid point $i$ and we write
\begin{equation}
C_{0i} = \frac{N_i \hat{B}_i}{4 \Delta\theta (\enr_{\rm max}-\enr_{\rm min}) \left[ u_{\rm max} - u_{\rm min} \right]_i}.
\label{eq:markers_c0_loc}
\end{equation}

\noindent The normalization constant is then obtained by averaging over all populated grid points:
\begin{equation}
C_0 = \left. \sum\limits_{i=1}^{N_{\rm g}} C_{0i} \right/ \sum\limits_{i=1}^{N_{\rm g}} \delta_{N_i > 0}; \qquad \text{where} \quad \delta_{N_i > 0} = \left\{\begin{array}{ccc} 1 & : & N_i > 0, \\ 0 & : & \text{else}. \end{array} \right.
\label{eq:markers_c0_avg}
\end{equation}

\noindent Alternatively, we could define
\begin{equation}
C_0\Delta\theta = \left(\sum_i N_i\right) \left/ \left\{ 4(\enr_{\rm max}-\enr_{\rm min})\sum_i[(u_{\rm max}-u_{\rm min})/\hat{B}]_i \right\} \right., \nonumber
\end{equation}

\noindent which gives a very similar value. However, the method of calculating local values $C_{0i}$ allows to check whether $C_{0i}$ is indeed approximately the same everywhere.

\subsubsection{Velocity space average}
\label{sec:num_df_dv}

The velocity space average of a quantity $X = X(\enr,\mu,\theta)$ is given by
\begin{equation}
\left< X \right>(\theta)
= 2\pi \sum\limits_{\sgnVp} \int\limits_0^\infty{\rm d}\enr \int\limits_0^{E/B}{\rm d}\mu\; \frac{B}{|v_\parallel|} X(\enr,\mu,\theta).
\end{equation}

\noindent Substituting the pitch angle variable $A = \mu B_0/\enr$ for $\mu$, and the normalized velocity $u=|v_\parallel|/\sqrt{2\enr}$ for $|v_\parallel|$, and sampling the phase space by including the factor $P_\delta/P_{\sgnVp}$ into the integrand yields
\begin{align}
\left< X \right>(\theta)
\approx & 2\pi \sum\limits_{\sgnVp} \int\limits_0^\infty{\rm d}\enr \int\limits_0^{1/\hat{B}}{\rm d}A\; \frac{\hat{B}\enr}{|v_\parallel|} \frac{P_\delta}{P_{\sgnVp}} X(\enr, A, \theta) \nonumber
\\
= & \frac{2\pi}{\sqrt{2}C_0} \sum\limits_{\sgnVp} \int\limits_0^\infty{\rm d}\enr \int\limits_0^{1/\hat{B}}{\rm d}A\; \hat{B} \sqrt{\enr} P_\delta X(\enr, A, \theta);
\end{align}

\noindent where $P_{\sgnVp} = C_0\sqrt{2\enr}/|v_\parallel|$ was used. Replacing the Dirac distribution $\delta(\theta - \theta_j)$ in Eq.~(\ref{eq:num_pdelta}) by the shape function $S(\theta - \theta_j)/\Delta\theta$, we obtain
\begin{equation}
\left< X \right>_i \approx
\frac{2\pi}{\sqrt{2}C_0\Delta\theta} \sum\limits_{j=1}^{N_{\rm m}} \hat{B}(\theta_j)\sqrt{\enr_j} S(\theta_i - \theta_j) X(\enr_j, A_j, \theta_j).
\label{eq:awecs_general_vspace_avg}
\end{equation}

\subsection{Boundary conditions}
\label{sec:num_bc}

The size of the simulation domain, $2\theta_{\rm max}$, depends largely on the magnetic shear $s$. The larger $s$ the faster any wave with finite radial extent is damped with increasing $|\theta|$. However, some eigenmodes, such as $\alpha$TAEs, couple to the Alfv\'{e}n continuum, which consists of waves singular in $r$. Since $r$ and $\theta$ are related though a Fourier transform, continuum waves in $\theta$-space take the form of undamped outgoing harmonic waves. In order to prevent unphysical reflections at the boundaries of the finite computational domain, the outgoing-wave boundary condition is facilitated by a ``boundary filter'' of the form
\begin{equation}
w(\theta) = \left\{
\begin{array}{lcl}
1 & : & |\theta| \leq \theta_{\rm bf}, \\
\exp\left(-\frac{|\theta| - \theta_{\rm bf}}{\sigma}\right) & : & |\theta| > \theta_{\rm bf}; \\
\end{array}
\right.
\end{equation}

\noindent which provides for artificial damping. Typically, $\theta_{\rm bf} = 0.8\times\theta_{\rm max}$ is used.

\subsection{Filtering}
\label{sec:num_filter}

We use the \textsc{fftw} package (version 2.1.5) for optional low-pass filtering in $k$-space. If enabled, the default cut-off is one quarter of the Nyquist wavenumber. In principle, this setting may reduce the accumulation of aliasing errors, but it cannot eliminate them. In linear gyrokinetic simulations, the amount of aliasing is mainly determined by the number of phase space markers and the shape function. To date, no significant effect of filtering on \textsc{awecs} simulation results has been observed, except for smoother mode structures.

\subsection{Algorithm}
\label{sec:num_algorithm}

The time integration can be carried out using either a 2nd- or a 4th-order Runge-Kutta scheme \cite{AbramowitzStegun}. The results presented in following sections were obtained with the 4th-order scheme. The computational cycle may be outlined as follows:
\begin{enumerate}
\item  Solve the GKE $\L_0 \delta g = -\delta\L \feqs \exp(i\delta_{{\rm d}s})$ along unperturbed marker orbits.

\item  Push markers along unperturbed particle orbits: $\theta_j(t) = \int{\rm d}t' v_{\parallel j}(\theta_j(t'))$.

\item  Calculate velocity space moments of the marker distribution.

\item  Evolve the electromagnetic fields $\dEe$ and $\dPsie$, given the moments of $\delta g$.

\item  Solve the algebraic field equations for $\dUe$ and $\dCe$, given $\dEe$, $\dPsie$ and the moments of $\delta g$.
\end{enumerate}

\noindent \textsc{awecs} automatically parallelizes on clusters using Message Passing Interface (MPI). A two-dimensional Cartesian processor grid is established in order to allow parallelization over markers and cases. Thus, \textsc{awecs} is capable of parallelizing case scans and producing organized output without the use of separate script files.

\section{Benchmark 1: Electrostatic instabilities}
\label{sec:bench_esitg}

\subsection{Preliminaries: ESITG equations}
\label{sec:bench_esitg_eqs}

In order to test the correct implementation of the solver algorithms for field and particle dynamics as well as the PIC method, a simple model for linear electrostatic ion-temperature-gradient-driven (ESITG) modes is implemented in \textsc{awecs-esitg}. After renormalizing the time as $\widetilde{t} = t\omega_{*{\rm i}}$, with $\omega_{*{\rm i}} = q \kiO \Ti^{1/2}/\eps_n$, and writing
\begin{equation}
\delta\widetilde{\phi} = -\tauei^T\dUe/\omega = e_{\rm i}\dphi/(\mi\Ti), \qquad
\delta h_{\rm i} = \delta g_{\rm i}/\omega, \nonumber
\end{equation}

\noindent we obtain the electrostatic limit ($\dPsie = \dCe = \Omega_p = 0$) by letting $\Ti \rightarrow 0$. Energetic particles are excluded and the resulting model equations are
\begin{align}
\delta\widetilde{\phi} =& \frac{\tauei^T}{1 + \tauei^T(1 - \Gamma_{0{\rm i}})} \left< J_{0{\rm i}} \delta h_{\rm i}\right>,
\label{eq:bench_esitg_dpsi}
\\
\frac{{\rm d}\theta_j}{{\rm d}\tilde{t}} =& \frac{\eps_n}{q\kiO} \hat{v}_\parallel,
\label{eq:bench_esitg_dx}
\\
\frac{{\rm d}\delta h_{\rm i}}{{\rm d}\tilde{t}} =& -i\tilde{\omega}_{\rm di} \delta h_{\rm i}
+ i\FMi \left[ (\tilde{\omega}_{*{\rm i}}^T - \tilde{\omega}_{\rm di}) J_{0{\rm i}} \delta\widetilde{\phi} + i\frac{\eps_n}{q\kiO} \hat{v}_\parallel \partial_\theta(J_{0{\rm i}} \delta\widetilde{\phi}) \right];
\label{eq:bench_esitg_dg}
\end{align}

\noindent where
\begin{equation}
\tilde{\omega}_{*{\rm i}}^T = 1 + \eta_{\rm i}\left( \hat{\enr} - 3/2 \right), \qquad
\tilde{\omega}_{\rm di} = \eps_n g \left(\hat{v}_\perp^2/2 + \hat{v}_\parallel^2\right), \qquad
\eps_n = \eps_{p{\rm i}} (1 + \eta_{\rm i}). \nonumber
\end{equation}

\noindent Note that, in the electrostatic limit, $\delta h_{\rm i} = \delta g_{\rm i} /\omega$ in Eq.~(\ref{eq:bench_esitg_dg}) is related to the perturbed distribution $\delta f_{\rm i}$ through the relation
\begin{equation}
\delta f_{\rm i} = -\frac{e_{\rm i}}{T_{\rm i}}\FMi(1 - J_{0{\rm i}}^2) \dphi + J_{0{\rm i}}\delta h_{\rm i}, \nonumber
\end{equation}

\noindent so it has a clear physical meaning: $J_{0{\rm i}} \delta h_{\rm i}$ is $\delta f_{\rm i}$ minus the density perturbation caused by the ion polarization drift.

\begin{figure}[tbp]
\includegraphics[width=1.0\textwidth]
{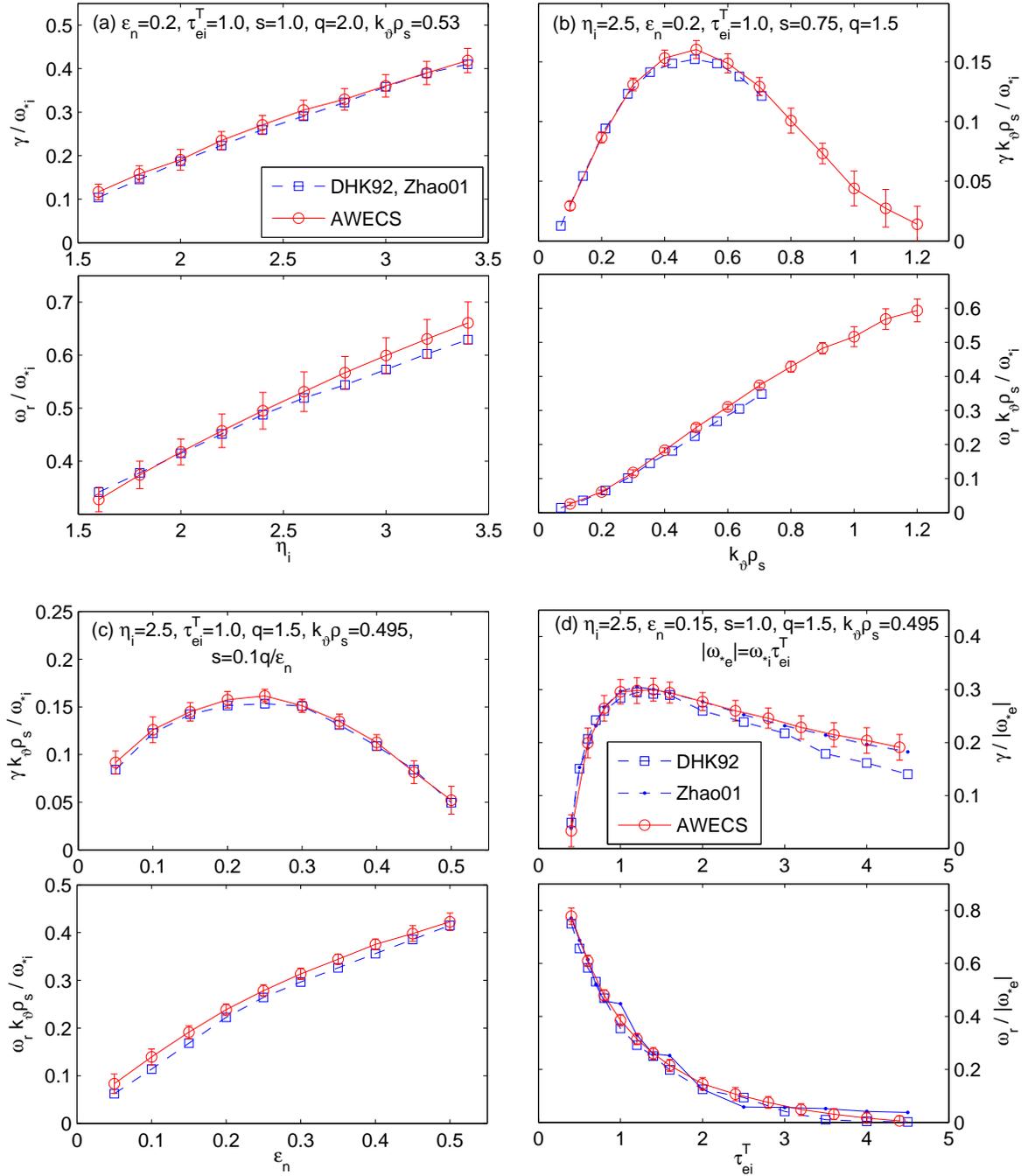}
\caption{Benchmark of \textsc{awecs} results for ESITG modes against Refs.~\protect\cite{Dong92, ZhaoThesis} (inverse aspect ratio $\eps = 0$). Except for the $\tauei^T$ scan (d), the results by Dong \textit{et al.}, 1992 (DHK92) \protect\cite{Dong92} and Zhao, 2001 (Zhao01) \protect\cite{ZhaoThesis} are very similar and are plotted as a single curve (squares).}
\label{fig:bench_esitg}%
\end{figure}

\subsection{Zero inverse aspect ratio}
\label{sec:bench_esitg_dong}

In order to test the correct implementation of passing particle dynamics, parameter scans with respect to $\eta_{\rm i}$, $k_\vartheta\rho_{\rm s} = (\tauei^T)^{1/2} \kiO$, $\varepsilon_n$ and $\tauei^T$ are carried out for cases with $\eps=0$, studied previously by Dong, Horton \& Kim, 1992 (DHK92) \cite{Dong92}. DHK92 solve an integro-differential formulation of the model as an eigenvalue problem. The resulting growth rates $\gamma$ and frequencies $\omega_{\rm r}$ are shown in Fig.~\ref{fig:bench_esitg}. Zhao, 2001 (Zhao01) \cite{ZhaoThesis} repeated DHK92's calculations with a code that solves Eqs.~(\ref{eq:bench_esitg_dpsi})--(\ref{eq:bench_esitg_dg}) as an initial value problem using a predictor-corrector scheme. Both calculations gave very similar results, so we have plotted them as a single curve for each case in Fig.~\ref{fig:bench_esitg}(a)--(c). A noticeable difference can be seen only in the $\tauei^T$ scan shown in Fig.~\ref{fig:bench_esitg}(d). The parameters used are
\begin{itemize}
\item  Physical parameters: $\varepsilon = \alpha = 0$. All other parameters are shown in Fig.~\ref{fig:bench_esitg}.

\item  Numerical parameters: $N_{\rm m} = 2048\times 4\times 3$, $\theta_{\rm max} = 20$, $N_{\rm g} = 256$, $\Delta t = 0.2$. In this and the following sections, we use $v_{\rm min} = 0.01$, $v_{\rm max} = 5.0$, $a_{\pi{\rm min}} = 0.0001$, $a_{\pi{\rm max}} = 0.9999$ [cf. Eq.~(\ref{eq:markers_input})], unless otherwise specified.
\end{itemize}

\noindent Note that DHK92 defines the thermal velocity as $\vti^{\rm Dong} = \sqrt{2\Ti}$, so we have scaled their input parameters and results by a factor $1/\sqrt{2}$ where appropriate ($k_\vartheta\rho_{\rm s}^{\rm Dong} = \sqrt{2} k_\vartheta\rho_s$). Furthermore, we have defined the diamagnetic frequency such that $\omega_{*{\rm i}} > 0$.

Figure~\ref{fig:bench_esitg} shows that the results by DHK92 and Zhao01 (squares) are reproduced accurately by \textsc{awecs-esitg} (circles). Runs with 2nd- and 4th-order Runge-Kutta schemes gave identical results.

\begin{figure}[tbp]
\includegraphics[width=1.0\textwidth]
{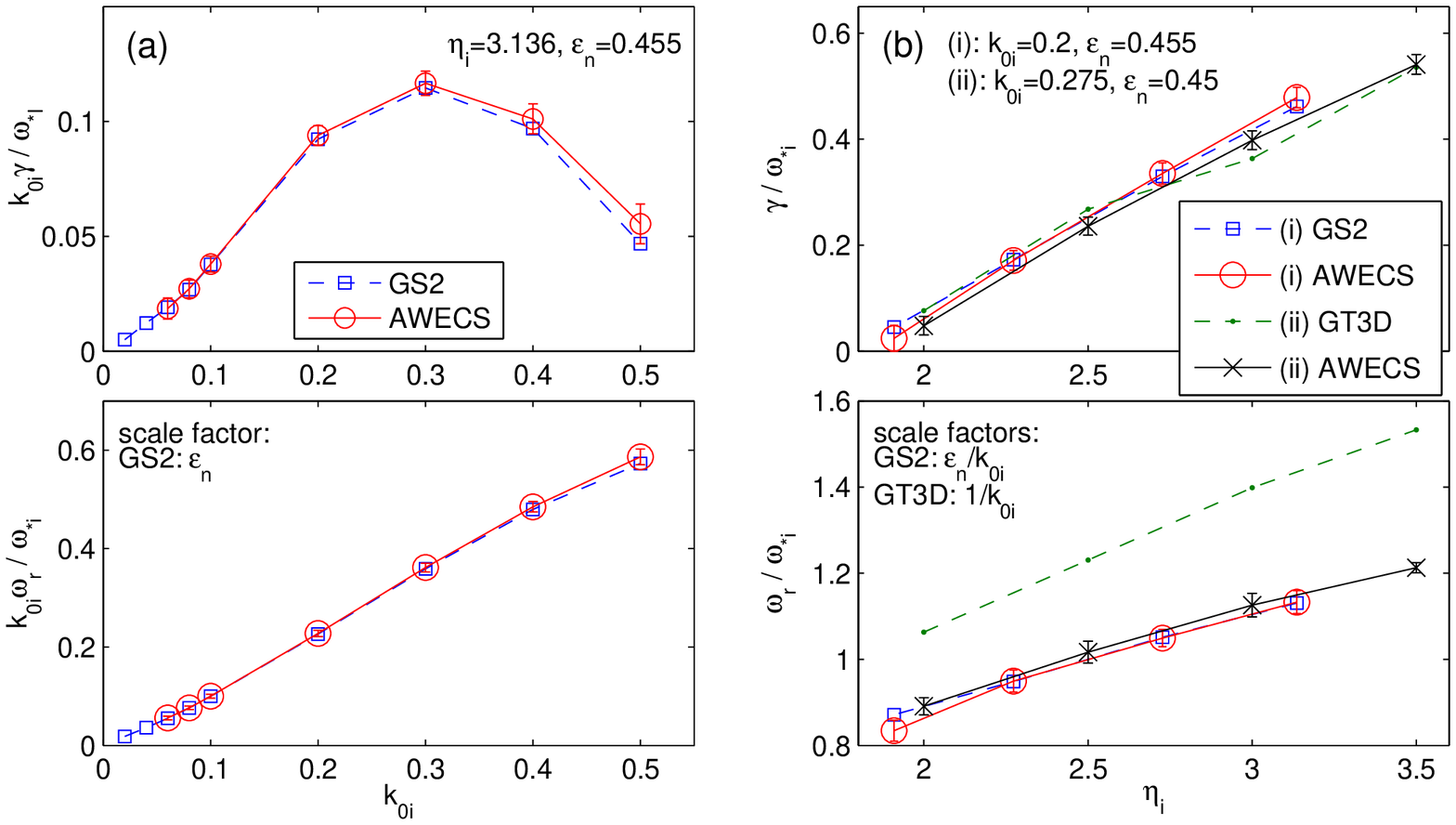}
\caption{Benchmark of \textsc{awecs} results for ESITG modes against Cyclone base case calculations performed with \textsc{gs2} \protect\cite{GS2_website} and \textsc{gt3d} \protect\cite{Idomura04} (inverse aspect ratio $\eps=0.18$).}
\label{fig:bench_esitg_cyclone}%
\end{figure}

\subsection{Cyclone base case}
\label{sec:bench_num_esitg_cyclone}

In order to test the correct implementation of trapped particle dynamics, we compare \textsc{awecs-esitg} results for the Cyclone base case with those produced by two other codes: an electrostatic version of the gyrokinetic fully-implicit initial-value code \textsc{gs2} \cite{GS2_website}, and the global gyrokinetic toroidal particle code \textsc{gt3d} \cite{Idomura04}. In the reference cases used, both \textsc{gs2} and \textsc{gt3d} were run in the massless-electron limit. Both \textsc{gs2} and \textsc{awecs} employ the $s$-$\alpha$ model and the main difference between the codes lies in the numerical scheme; thus, the results should be quantitatively comparable. When comparing with the global code \textsc{gt3d}, we only look for qualitative similarity between the results. The parameters are:

\begin{itemize}
\item  Physical parameters: The Cyclone base case parameters are $\eps = 0.18$, $\eta_{\rm i} = 3.114$, $\eps_n = 0.45$, $s = 0.776...0.796$, $q = 1.4$, $\tauei^T = 1.0$ \cite{Dimits00}. The actual input parameters used in \textsc{gs2} and \textsc{gt3d} differ slightly and we use their settings for our calculations. The parameters to be varied are $\kiO$ and $\eta_{\rm i}$.

\item  Numerical parameters: $N_{\rm m} = 2048\times 4\times 3$, $\theta_{\rm max} = 20$, $N_{\rm g} = 256$, $\Delta t = 0.2$. The bounce angle range is $\theta_{\rm b,min} = 0.1^\circ$ and $\theta_{\rm b,max} = 179.9^\circ$.
\end{itemize}

Figure~\ref{fig:bench_esitg_cyclone}(a) shows ESITG growth rates $\gamma$ and frequencies $\omega_{\rm r}$ in dependence of $\kiO$. The results show good agreement between \textsc{gs2} and \textsc{awecs}. Figure~\ref{fig:bench_esitg_cyclone}(b) shows results of an $\eta_{\rm i}$ scan. Both growth rates and frequencies agree with \textsc{gs2} results. The comparison with \textsc{gt3d} shows surprisingly good quantitative agreement in the growth rates $\gamma$. The systematic discrepancy of about $20\%$ in the frequencies $\omega_{\rm r}$ may be due to the fact that \textsc{awecs} is a local code and uses the CHT $s$-$\alpha$ equilibrium, whereas \textsc{gt3d} is a global code.

\section{Benchmark 2: Shear-Alfv\'{e}n instabilities}
\label{sec:bench_saw}

\subsection{Preliminaries: MHD SAW equation and terminology}
\label{sec:model_saw_eq}

Let us first consider shear Alfv\'{e}n waves with $\omega \sim \omAO$, ignore the temperature gradient ($\eta_s = 0$), and treat the plasma in the ideal-MHD limit ($\dUe = 0$). In the cold-ion limit ($\omega_{\rm r} \gg \{k_\parallel v_\parallel, \omega_{*{\rm i}}, \omegadi\}$, $\nEO = 0$ and $\kiO \ll 1$), the contributions from kinetic terms and $\dCe$ can be neglected and the field equations reduce to the low-$\beta$ ideal-MHD SAW equation
\begin{equation}
\partial_\theta (f \partial_\theta \dPsie) + (1 + 2\eps\cos\theta) f \omega^2 \dPsie + \alpha g\dPsie = 0.
\label{eq:model_saw_cold}
\end{equation}

\noindent FLR effects may be taken into account by retaining the $\omega_{*p{\rm i}}$ correction to the inertia term, which yields an MHD SAW equation applicable to a warm plasma \cite{Tang81}. In this case, $\omega^2$ in Eq.~(\ref{eq:model_saw_cold}) is replaced by $\omega(\omega - \omega_{*p{\rm i}})$. The substitutions $\delta\Psi_{\rm s} = \sqrt{f} \dPsie$ and $\Omega = \omega - \omega_{*p{\rm i}}/2$ turn the SAW equation into Schr\"{o}dinger-like form,
\begin{equation}
\partial_\theta^2\delta\Psi_{\rm s} + (1 + 2\eps\cos\theta) (\Omega^2 - \Omega_*^2) \delta\Psi_{\rm s} - V_{\rm ball} \delta\Psi_{\rm s} = 0;
\label{eq:model_saw_eq}
\end{equation}

\noindent where
\begin{equation}
V_{\rm ball} = (s-\alpha\cos\theta)^2/f^2 - \alpha\cos\theta/f \nonumber
\end{equation}

\noindent is the ballooning potential and $\Omega_*^2 = \omega_{*p{\rm i}}^2/4$ is a constant offset. Equation~(\ref{eq:model_saw_eq}) describes the propagation of periodically driven waves ($\eps\cos\theta$ factor) in the potential $V_{\rm ball}$. Using an MHD shooting code, we solve Eq.~(\ref{eq:model_saw_eq}) to obtain MHD SAW frequencies in the high-$\beta$ regime, which will be useful to verify results of gyrokinetic simulations.

A derivation similar to that giving Eq.~(\ref{eq:model_saw_cold}), but for marginally stable modes ($\gamma = 0$) with frequencies $\omega_{\rm r} = \omega_{*{\rm i}}$ (again letting $\eta_{\rm i} = 0$), yields the Connor-Hastie-Taylor (CHT) MHD ballooning equation \cite{Connor78}
\begin{equation}
\partial_\theta^2\delta\Psi_{\rm s} - V_{\rm ball} \delta\Psi_{\rm s} = 0.
\label{eq:model_saw_ball}
\end{equation}

\begin{figure}[tb]
\includegraphics[width=1.00\textwidth]
{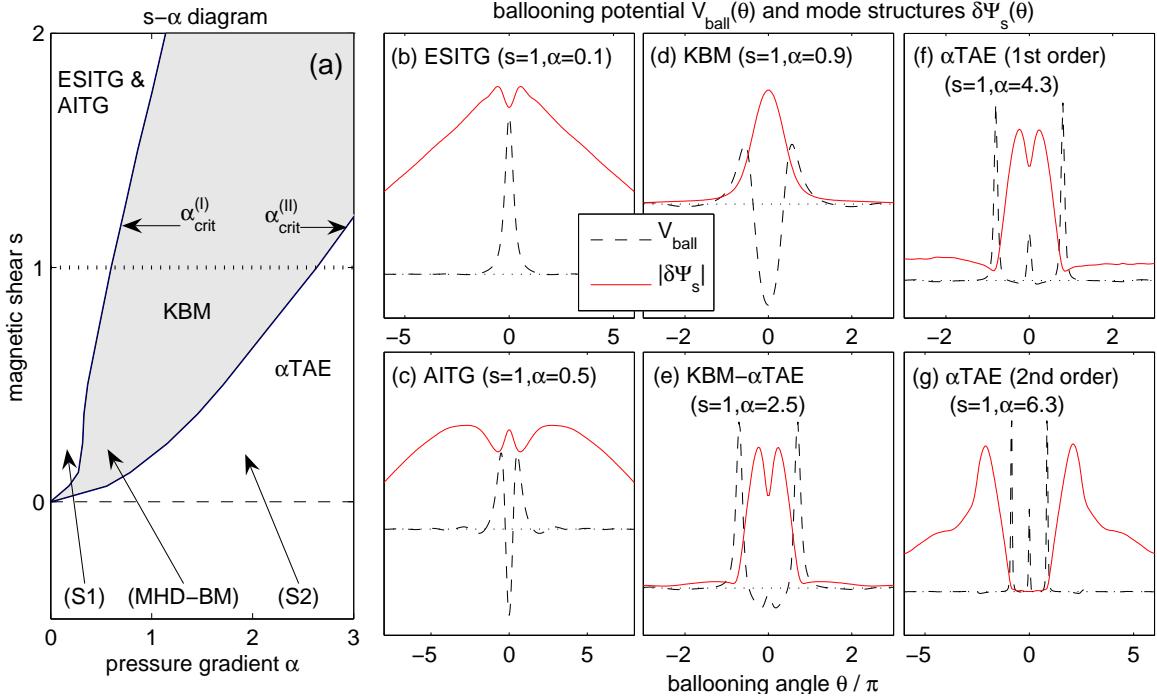}
\caption{(a): $s$-$\alpha$ diagram showing (S1) is the first stable domain, (S2) the second stable domain, and (MHD-BM) the high-$n$ MHD ballooning unstable domain for $\theta_k = 0$. (b)--(g): Ballooning potential $V_{\rm ball}(\theta|s,\alpha)$ and mode structures of instabilities used for benchmarking. See also Table~\protect\ref{tab:bench_saw}.}
\label{fig:bench_saw_salpha}%
\end{figure}

\noindent Equation~(\ref{eq:model_saw_ball}) determines the stability boundaries of ideal MHD ballooning modes as shown in the $s$-$\alpha$ diagram in Fig~\ref{fig:bench_saw_salpha}(a). Ballooning modes (BM) are short-wavelength (high-$n$) pressure-gradient-driven instabilities similar to the Rayleigh-Taylor interchange instability. They are localized in regions where the field line curvature ${\bm \kappa}$ is unfavorable; i.e., where ${\bm \kappa}$ has the same sign as the pressure gradient. The $s$-$\alpha$ diagram in Fig~\ref{fig:bench_saw_salpha}(a) shows the stability boundaries $\acriti(s)$ and $\acritii(s)$, which divide the plane into the first stable domain (S1), the second stable domain (S2), and the MHD ballooning unstable domain (MHD-BM). The diagram shows the case $\theta_k = 0$; the boundaries are modified for finite $\theta_k$ \cite{Chen87}, which is not considered here. The Mercier criterion, which determines the minimum values of $s$ and $\alpha$ for interchange instability to occur, is not contained in the CHT $s$-$\alpha$ model, so that the (MHD-BM) domain in Fig~\ref{fig:bench_saw_salpha}(a) extends to $s=\alpha=0$.

\begin{table}[bt]
\centering
\begin{tabular}{ccccc}
\hline\hline
$\alpha \ll \acriti$ & $\alpha \lesssim \acriti$ & $\acriti \lesssim \alpha \lesssim \acritii$ & $\alpha \gtrsim \acritii$ \\
\hline
ESITG & AITG & KBM & $\alpha$TAE \\
\hline\hline
\end{tabular}
\caption{Instabilities and MHD eigenmodes observed in the benchmark simulations. ESITG: electrostatic ion-temperature-gradient driven mode; AITG: Alfv\'{e}nic ITG mode; KBM: kinetic ballooning mode; $\alpha$TAE: $\alpha$-induced toroidal Alfv\'{e}n eigenmode. See Fig.~\protect\ref{fig:bench_saw_salpha}(b)--(g) for corresponding mode structures.}
\label{tab:bench_saw}
\end{table}

When kinetic particle compression is taken into account, both stable domains (S1) and (S2) become populated with temperature-gradient- and pressure-gradient-driven kinetic instabilities. \textsc{awecs} has been developed to study these modes. In the following, we will consider several previously studied cases to verify that \textsc{awecs} reproduces earlier results correctly. The instabilities considered are listed in Table~\ref{tab:bench_saw} and examples for the respective mode structures are shown in Fig.~\ref{fig:bench_saw_salpha}(b)--(g).

The instability shown in Fig.~\ref{fig:bench_saw_salpha}(g) was discovered by Hirose \textit{et al.} \cite{Hirose94}. The mode structure peaks in regions of unfavorable curvature outside the central potential well, so this kind of instability was called ``higher-order ballooning mode.'' Based on the fact that the frequency $\omega_{\rm r}$ approaches a nonzero value as $\kiO \rightarrow 0$ (cf.~Fig.~15 in Ref.~\protect\cite{Hirose95}), we conclude that these modes are closely related to discrete MHD Alfv\'{e}n eigenmodes known to exist in the second MHD ballooning stable domain \cite{Hu04}. These are called $\alpha$TAE, so this name is used in the present work. The names ``ballooning mode'' and ``AITG mode'' are reserved for instabilities with, in the incompressible thermal ion limit, $\omega_{\rm r}(\kiO\rightarrow 0) \rightarrow 0$ residing inside and outside the (MHD-BM) domain, respectively.

In this section, we benchmark \textsc{awecs} for shear-Alfv\'{e}n instabilities in the absence of energetic particles. Benchmarks including energetic particles are presented Section~\ref{sec:bench_atae}.

\subsection{Ballooning stability boundaries}
\label{sec:bench_saw_ball}

\subsubsection{Ideal-MHD limit}
\label{sec:bench_saw_ball_mhd}

\begin{figure}[tbp]
\includegraphics[width=1.00\textwidth]
{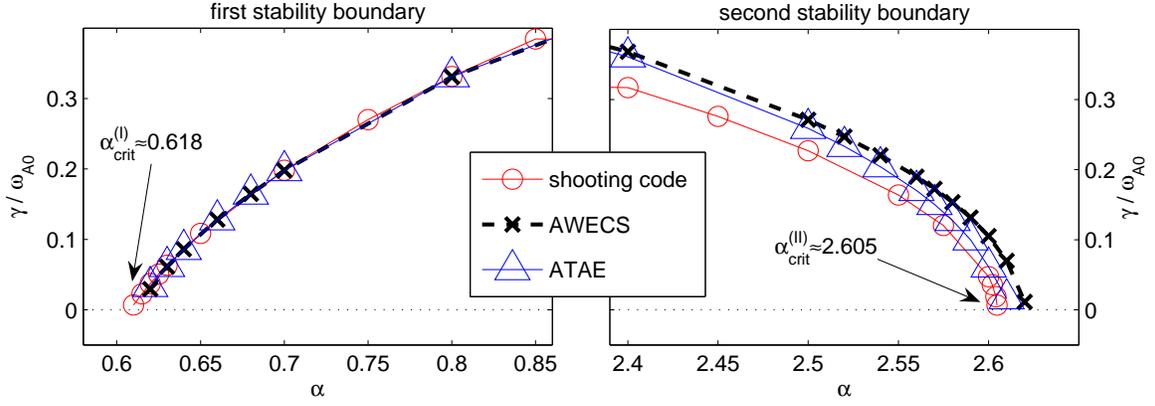}
\caption{Benchmark of ideal-MHD ballooning stability boundaries. The growth rate $\gamma$ is plotted as a function of $\alpha$, (a) near $\acriti$ and (b) near $\acritii$ for $s=1.0$.}
\label{fig:bench_saw_ball_mhd}%
\end{figure}

In Fig.~\ref{fig:bench_saw_ball_mhd} we compare results results obtained with a shooting code and two initial value codes, \textsc{awecs} (4th-order Runge-Kutta) and \textsc{atae} (2nd-order leap frog). Both \textsc{atae} and the shooting code solve Eq.~(\ref{eq:model_saw_cold}). The size of the computational domain in the initial value codes is $\theta_{\rm max} \approx 6\pi...17\pi$, whereas the shooting code is run with $\theta_{\rm max}$ up to $256\pi$. Thus, the latter is expected to give the most accurate results for $\acriti$ or $\acritii$, since the mode structure becomes very broad in $\theta$ when $\alpha$ approaches marginal stability. The parameters used in \textsc{awecs} are: $q = 1.5$, $s=1.0$, $\eps_n=0.2$, $\kiO=0.001$, $\tauei^T = 1$, $\eta_s = 0$. The effects of $\dBp$ and $\Omega_p$ are negligible. Kinetic terms are turned off by setting $\delta G_{\rm i} = 0$.

Results of scans of the parameter $\alpha$ near the first and second ballooning stability boundary are shown in Fig.~\ref{fig:bench_saw_ball_mhd}, and the values obtained for $\acriti$ and $\acritii$ are summarized in Table~\ref{tab:bench_saw_ball_mhd}. From the good agreement between the three codes it can be concluded that \textsc{awecs} reproduces the ideal-MHD stability boundaries correctly.

\begin{table}[btp]
\centering
\begin{tabular}{cccc}
\hline\hline
& shooting code & \textsc{atae} & \textsc{awecs} \\
\hline
$\acriti$: & 0.608 & 0.61 & 0.61 \\
$\acritii$: & 2.605 & 2.61 & 2.62 \\
\hline\hline
\end{tabular}
\caption{Benchmark of MHD ballooning stability boundaries.}
\label{tab:bench_saw_ball_mhd}
\end{table}

\subsubsection{Finite ion Larmor radius and diamagnetic drift frequency}
\label{sec:bench_saw_ball_gk}

It can be shown that for $\eta_{\rm i}=0$, all FLR terms cancel for modes satisfying $\omega \approx \omega_{*{\rm i}} = \omega_{*p{\rm i}}$, and the CHT ballooning equation (\ref{eq:model_saw_ball}) is valid for any value of $\kiO$ \cite{Zhao02}. On the other hand, for the general case $\omega_{\rm r} \neq \omega_{*p{\rm i}}$, FLR effects were shown to have a stabilizing effect on ballooning modes, yielding somewhat higher values for $\acriti$ \cite{Tang81}.

\begin{figure}[tbp]
\includegraphics[width=1.00\textwidth]
{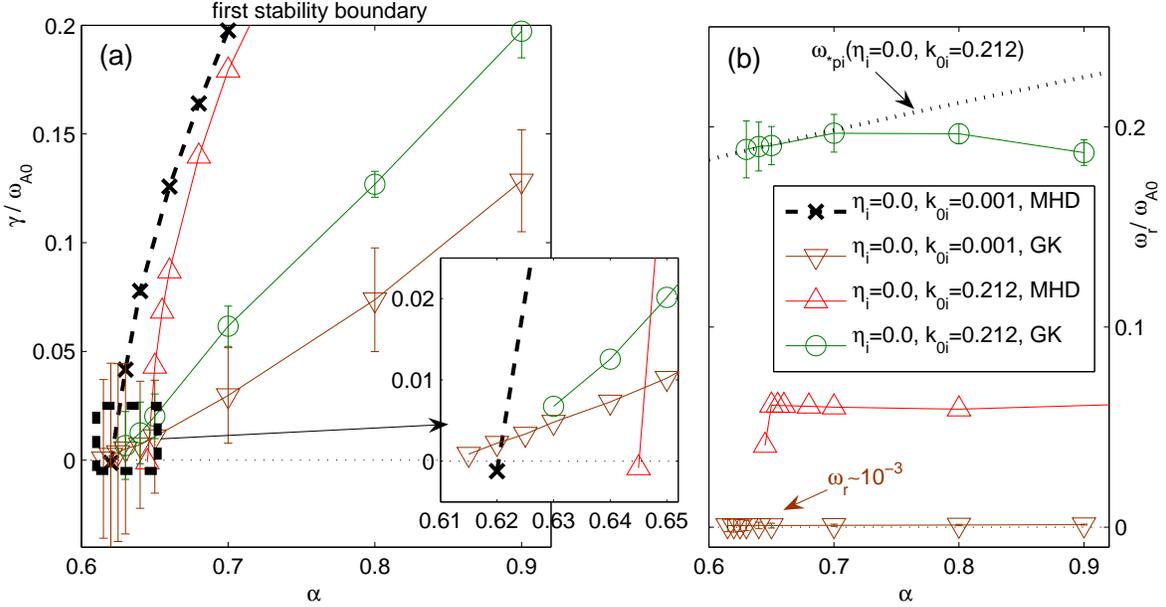}
\caption{Qualitative behavior of ballooning stability boundaries for finite $\kiO$ with kinetic terms turned on (GK) and off (MHD). The growth rate $\gamma$ (a) and frequency $\omega_{\rm r}$ (b) are plotted as a functions of $\alpha$ near $\acriti$ for $s=1.0$.}
\label{fig:bench_saw_ball_gk}%
\end{figure}

To verify that \textsc{awecs} reproduces this qualitative behavior, we repeat the calculation of Section~\ref{sec:bench_saw_ball_mhd} with $\kiO = 0.001$ and $\kiO = 0.212$, while still setting $\eta_s = 0$. Each of these two cases is calculated once with the kinetic terms turned off (``MHD'') and once turned on (``GK''). The results of an $\alpha$-scan near the first stability boundary $\acriti$ are shown in Fig.~\ref{fig:bench_saw_ball_gk}.

In the case (MHD, $\kiO = 0.001$), the ideal MHD stability boundary is reproduced with good accuracy: $\acriti \approx 0.615$. In the case (MHD, $\kiO = 0.212$), a somewhat larger value $\acriti \approx 0.64$ is obtained, and the mode has a frequency $\omega_{\rm r} \sim \omega_{*{\rm i}}/3$ near the stability boundary.

Let us now consider the two cases where the kinetic compression terms are retained (``GK''). Extrapolation of the averaged growth rates suggests that the modes become stable near the ideal MHD value $\alpha_{\rm crit}^{\rm (I)} \approx 0.615$. However, the signals are oscillating which gives rise to the large error bars shown in Fig.~\ref{fig:bench_saw_ball_gk}(a) and (b). The oscillation amplitude is larger for $\kiO = 0.001$ than for $\kiO = 0.212$.\footnote{In Section~\protect\ref{sec:bench_saw_lowb}, we will show that the growth rates in the latter case, (GK, $\kiO = 0.212$), compare well with the results obtained by Zhao \& Chen \cite{Zhao02}.} Note that, in contrast to the case (MHD, $\kiO = 0.212$), the frequency in case (GK, $\kiO = 0.212$) approaches $\omega_{*p{\rm i}}$ near the stability boundary. However, this should be considered coincidental; the two frequencies happen to be the same for this particular value of the wavenumber, $\kiO = 0.212$, while they differ for different values of $\kiO$. Yet, for this particular case, the extrapolated $\acriti$ coincides with the MHD value, so the result is consistent with the requirement that $\acriti$ be independent of $\kiO$ if $\omega \sim \omega_{*{\rm i}} = \omega_{*p{\rm i}}(\eta_{\rm i}=0)$.

It can be concluded that \textsc{awecs} correctly reproduces the qualitative behavior of the first ballooning stability boundary for finite values of the ion Larmor radius.

\subsection{Low-$\bm\beta$ instabilities}
\label{sec:bench_saw_lowb}

The kinetic excitation of electrostatic and Alfv\'{e}nic instabilities in the first ballooning stable domain and near $\acriti$ has been studied by Dong, Chen \& Zonca, 1999 (DCZ99) using an eigenvalue solver \cite{Dong99}, and by Zhao \& Chen, 2002 (ZC02) using an initial value code \cite{Zhao02}. In this section, we compare \textsc{awecs} results with data obtained in these two earlier studies. The parameters used are
\begin{itemize}
\item  Physical parameters: $q=1.5$, $s=1.0$, $\eps_n=0.2$, $\kiO = 0.212$ ($k_\vartheta\rho_{\rm s}^{\rm Dong} = \sqrt{2}\kiO = 0.3$), $\tauei^T = 1.0$, $\eta_{\rm i} = \eta_{\rm e}$, $\eps = 0$.

\item  Numerical parameters: $N_{\rm m} = 512\times 4\times 5$, $\theta_{\rm max} = 40$, $N_{\rm g} = 512$, $\Delta t = 0.02$.
\end{itemize}

\noindent \textsc{awecs} was run with and without magnetic compression and identical results were obtained, which justifies the approximation $\dBp = \Omega_p = 0$ used in Refs.~\cite{Dong99, Zhao02}.

\begin{figure}[tbp]
\includegraphics[width=1.00\textwidth]
{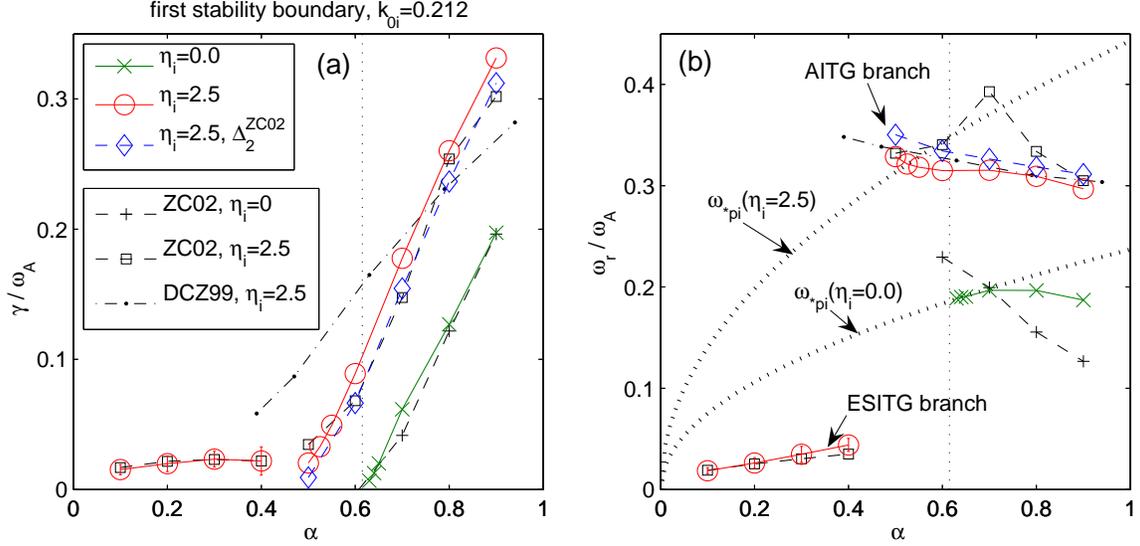}
\caption{Benchmark in the first ballooning stable domain and near $\acriti$. Growth rates (a) and frequencies (b) obtained using \textsc{awecs} are compared with results obtained by Dong \textit{et al.}, 1999 (DCZ99) \protect\cite{Dong99} and Zhao \& Chen, 2002 (ZC02) \protect\cite{Zhao02}. Two cases are shown: $\eta_{\rm i} = 0$ and $\eta_{\rm i} = 2.5$.}
\label{fig:bench_saw_zhao}%
\end{figure}

The results are shown in Fig.~\ref{fig:bench_saw_zhao}. The growth rates in both cases, $\eta_{\rm i} = 0$ and $\eta_{\rm i} = 2.5$, agree well with ZC02's results\footnote{The growth rates in Ref.~\protect\cite{Zhao02} seem to be incorrect; a factor of $\sqrt{2}$ had to be applied in order to obtain agreement with \textsc{awecs} results. The authors of Ref.~\protect\cite{Zhao02} have scaled the results by Dong \textit{et al.} \protect\cite{Dong99} by $\sqrt{2}$ for comparability and may have accidentally scaled their own growth rate data as well. It is not clear whether the frequencies are also affected; in Fig.~\protect\ref{fig:bench_saw_zhao}, the frequency values from Ref.~\protect\cite{Zhao02} are plotted without additional scaling, because the frequencies of the ESITG branch coincide as they are.}
and give the same critical $\alpha$ values: $\acriti(\eta_{\rm i}=0) \approx 0.61$ (KBM) and $\acriti(\eta_{\rm i}=2.5) \approx 0.5$ (AITG). The small discrepancy in the growth rates of the AITG branch is due to the fact that our calculation yields a different coefficient $\Delta_2$ than that used by ZC02 [see Footnote before Eq.~(\ref{eq:model_final_dh1})]. If $\Delta_2^{\rm ZC02}$ is adopted, \textsc{awecs} reproduces their result exactly as can be seen in Fig.~\protect\ref{fig:bench_saw_zhao}. The AITG frequencies obtained by ZC02 agree with \textsc{awecs} results, except for one point: $\alpha=0.75$. The reason for this deviation is not known. As mentioned in the previous section, the fact that the frequency $\omega_{\rm r}$ approaches $\omega_{*p{\rm i}}$ near the stability boundary is coincidental and depends on the value of $\kiO$.

The AITG growth rates by DCZ99 differ from \textsc{awecs} results and those obtained by ZC02. On the other hand, the frequencies agree well with those obtained using \textsc{awecs}. The reason for the discrepancy in the growth rates is not known. ZC02 speculated that the problem may be due to the stronger sensitivity of the eigenvalue code to boundary effects. However, the mode structure is broad only near the stability boundary, while the discrepancy persists for $\alpha > \acriti$ (see also Section~\ref{sec:bench_saw_highb}). Moreover, Dong \textit{et al.}, 2004 (DCZJ04) \cite{Dong04} revisited this calculation with $\theta_{\rm max} \approx 23$ and reproduced the earlier results by DCZ99, where $\theta_{\rm max} \approx 16$ was used. On the other hand, our convergence study in Section~\ref{sec:accurate_converge} will show that $\theta_{\rm max} \gtrsim 40$ may be necessary.

Note that the ESITG branch ($\eta_{\rm i}=2.5$, $\alpha < 0.5$) is reproduced accurately; both the frequencies and the growth rates agree well. ESITG results are not affected by the different $\Delta_2$ used by ZC02.

Based on the agreement with results by ZC02 (major discrepancies were explained), we will assume that \textsc{awecs} correctly computes the properties of finite-$\beta$ ESITG (drift-Alfv\'{e}n waves), AITG modes and KBMs near the first stability boundary. Discrepancies between the initial value code results and those obtained with the eigenvalue approach used by DCZ99 and DCZJ04 remain to be understood.

\begin{figure}[tbp]
\includegraphics[width=0.95\textwidth]
{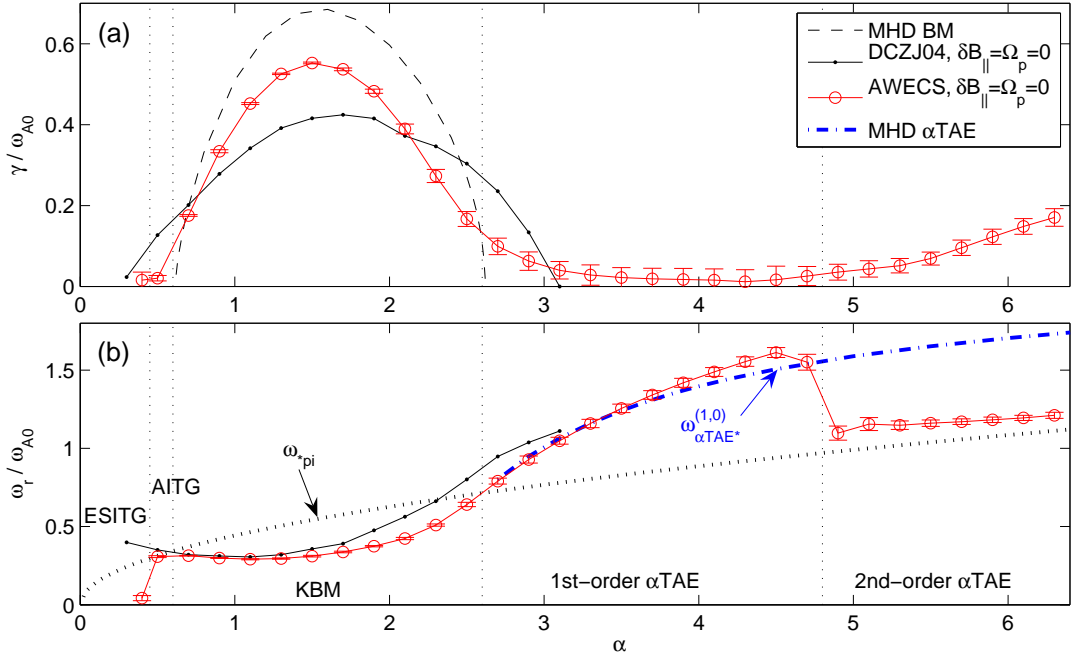}
\caption{Benchmark for $s=1.0$ over a wide range of $\alpha$ values, including the second ballooning stable domain. Growth rates $\gamma$ (a) and frequencies $\omega_{\rm r}$ (b) obtained using \textsc{awecs} (solid line with circles) are compared with results obtained by Dong \textit{et al.}, 2004 (DCZJ04) \protect\cite{Dong04} (solid line with dots). In addition, the MHD frequencies for 1st-order $\alpha$TAE modes (with $\omega_{*p{\rm i}}$ corrections) are plotted (dash-dotted line).}
\label{fig:bench_saw_dong}%
\end{figure}

\begin{figure}[tb]
\includegraphics[width=0.95\textwidth]
{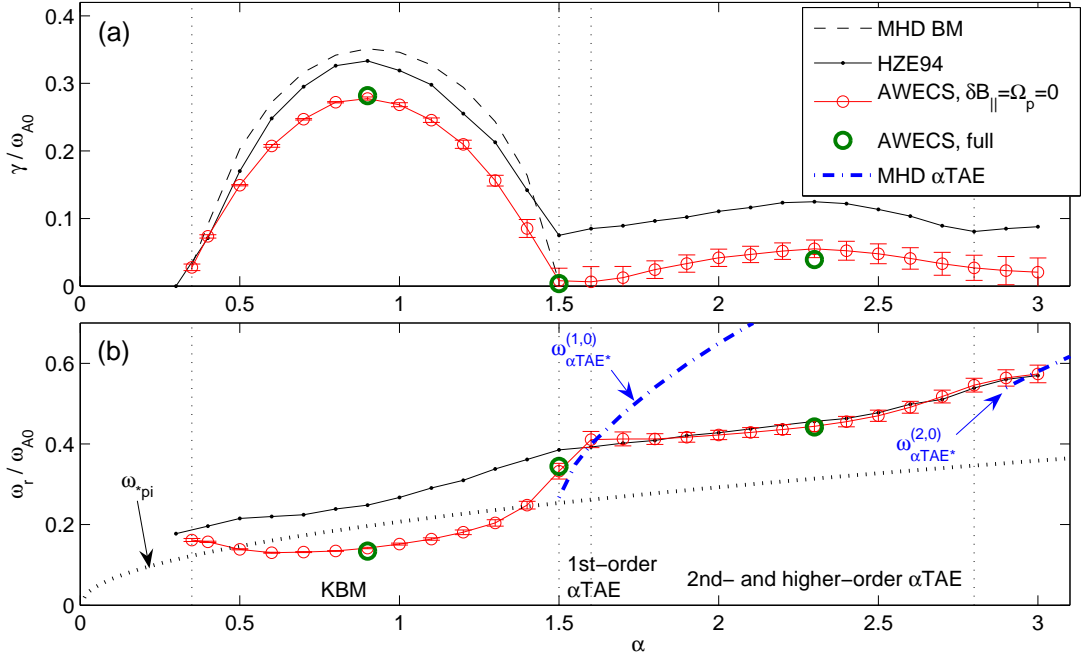}
\caption{Benchmark for $s=0.4$ over a wide range of $\alpha$ values, including the second MHD ballooning stable domain. Growth rates $\gamma$ (a) and frequencies $\omega_{\rm r}$ (b) obtained using \textsc{awecs} are compared with results obtained by Hirose \textit{et al.}, 1994 (HZE94) \protect\cite{Hirose94}. In (b), the frequencies of the primary unstable mode from HZE94 is plotted. In addition, the MHD frequencies for 1st- and 2nd-order $\alpha$TAE modes (with $\omega_{*p{\rm i}}$ corrections) are plotted (dash-dotted lines).}
\label{fig:bench_saw_hirose}%
\end{figure}

\subsection{High-$\bm\beta$ instabilities}
\label{sec:bench_saw_highb}

The kinetic excitation of Alfv\'{e}nic instabilities near $\acritii$ and in the second ballooning stable domain has been studied by Dong \textit{et al.}, 2004 (DCZJ04) using an eigenvalue solver \cite{Dong04}, and by Hirose, Zhang \& Elia, 1994 (HZE94) using a shooting code \cite{Hirose94}. In this section, we compare \textsc{awecs} results with results obtained in these two earlier studies.

We begin with a discussion of a case studied by DCZJ04, shown in Fig.~\ref{fig:bench_saw_dong}, where the following parameters were used
\begin{itemize}
\item  Physical parameters: $q=1.5$, $s=1.0$, $\eps_n=0.2$, $\kiO = 0.212$ ($k_\vartheta\rho_{\rm s}^{\rm Dong} = \sqrt{2}\kiO = 0.3$), $\tauei^T = 1.0$, $\eta_s = 2.5$, $\eps = 0$.

\item  Numerical parameters: $N_{\rm m} = 512\times 4\times 7$, $\theta_{\rm max} = 40$, $N_{\rm g} = 512$, $\Delta t = 0.01$.
\end{itemize}

\noindent In the domain scanned by DCZJ04, $0.3 \leq \alpha \leq 3.1$, there is a significant difference in the growth rate and critical $\alpha$ values. In fact, DCZJ04 reported that no instability was observed for $\alpha > 3.1$; which does not agree with our findings. Although, we are not able to verify the growth rates (except for a convergence study), we can confirm that the instability observed in our results is a physical mode: The frequency in the region $2.8 \lesssim \alpha \lesssim 4$ follows closely the frequency of the first-order $\alpha$TAE obtained with the MHD shooting code, which is denoted by $\omega_{\alpha{\rm TAE}*}^{(1,0)}$ in Fig.~\ref{fig:bench_saw_dong} (dash-dotted line). The corresponding mode structure is plotted in Fig.~\ref{fig:bench_saw_salpha}(f).

Here and in the following, $\omega_{\alpha{\rm TAE}*}^{(j,p)}$ denotes the $\omega_{*{\rm pi}}$-corrected MHD frequency of an $\alpha$TAE. The ``order'' $j\geq 1$ identifies in which potential well the mode amplitude has its maximum, and $p\geq 0$ counts the number of zeros the mode structure has in potential well $j$ \cite{Hu04, Hu05}. The ground state corresponds to $p=0$. The frequencies $\omega_{\alpha{\rm TAE}*}^{(j,p)}$ are calculated by a MHD shooting code solving Eq.~(\ref{eq:model_saw_eq}). In this work, only solutions obtained by shooting along the real $\theta$ axis are considered. A more complete study would require shooting into the complex plane and the application of phase-integral methods \cite{Hu05}, since $\alpha$TAEs are generally coupled to the Alfv\'{e}n continuum, either directly or via barrier tunneling. This implies that the problem requires an outgoing-wave boundary condition and that, in general, the solutions of interest are not square integrable. This may explain why the eigenvalue solver DCZJ04 yields a different result.

For $\alpha > 4.8$, a 2nd-order $\alpha$TAE, with frequency $\omega_{\rm r} < \omega_{\alpha{\rm TAE}*}^{(1,0)}$, becomes the dominant instability. The corresponding mode structure is plotted in Fig.~\ref{fig:bench_saw_salpha}(g). In the MHD limit, this mode is strongly damped in the entire $\alpha$-range scanned, so no shooting result for $\omega_{\alpha{\rm TAE}*}^{(2,0)}$ is available. However, a $\kiO$-scan at $\alpha = 6.0$ revealed that $\omega_{\rm r}(\kiO\rightarrow 0)$ is nonzero, which supports our assertion that the mode is an $\alpha$TAE.

Let us now proceed to the case studied by HZE94, shown in Fig.~\ref{fig:bench_saw_hirose}, where the following parameters were used
\begin{itemize}
\item  Physical parameters: $q=1.2$, $s=0.4$, $\eps_n=0.175$, $\kiO = 0.1$, $\tauei^T = 1.0$, $\eta_{\rm i} = \eta_{\rm e} = 2.0$, $\eps = 0$.

\item  Numerical parameters: $N_{\rm m} = 512\times 4\times 7$, $\theta_{\rm max} = 60$, $N_{\rm g} = 1024$, $\Delta t = 0.02$.
\end{itemize}

\noindent It must be noted that the term $v_\parallel\partial_\theta$ is omitted in the model used by HZE94, so that important physical effects associated with the transit resonance are missing.

Figure~\ref{fig:bench_saw_hirose}(a) shows that the qualitative behavior of the growth rate seen by HZE94 is reproduced by \textsc{awecs}, both for KBM and $\alpha$TAE. The fact that the growth rate in HZE94 are systematically larger is most likely related to their neglect of the transit resonance (Landau damping). Note that the transition to a 3rd-order mode around $\alpha \approx 2.8$ apparent in HZE94's results is not obvious in the \textsc{awecs} data. For more accurate simulations in this regime, markers must be loaded in a broader range of the simulation domain. The size of the simulation domain itself may also need to be increased.

Near the second MHD ballooning stability boundary, in the range $1.5 \lesssim \alpha \lesssim 1.6$, the mode frequency $\omega_{\rm r}$ is similar to that of the 1st-order $\alpha$TAE obtained with the MHD shooting code, $\omega_{\alpha{\rm TAE}*}^{(1,0)}$, as can be seen in Fig.~\ref{fig:bench_saw_hirose}(b). In the range $1.6 \lesssim \alpha \lesssim 3$, a 2nd-order $\alpha$TAE is excited. This mode is strongly damped for $\alpha \lesssim 2.8$, so shooting results for $\omega_{\alpha{\rm TAE}*}^{(2,0)}$ are available only for $\alpha > 2.8$. However, a $\kiO$-scan at $\alpha = 2.3$ revealed that $\omega_{\rm r}(\kiO\rightarrow 0)$ is nonzero, which supports our assertion that the mode is an $\alpha$TAE.

In summary, the qualitative agreement between \textsc{awecs} results and those by HZE94, and the agreement of the frequencies with those obtained for MHD $\alpha$TAEs suggests that \textsc{awecs} accurately reproduces essential properties of high-$\beta$ shear-Alfv\'{e}n instabilities. Discrepancies with results by DZCJ04 remain to be resolved. Further simulations using \textsc{awecs} indicate that the effects of $\dBp$ and $\Omega_p$ are small in the $\alpha$ range scanned in the two cases discussed above. As an example, several points obtained using the full code are plotted in Fig.~\ref{fig:bench_saw_hirose} (bold circles).

\begin{figure}[tb]
\includegraphics[width=1.0\textwidth]
{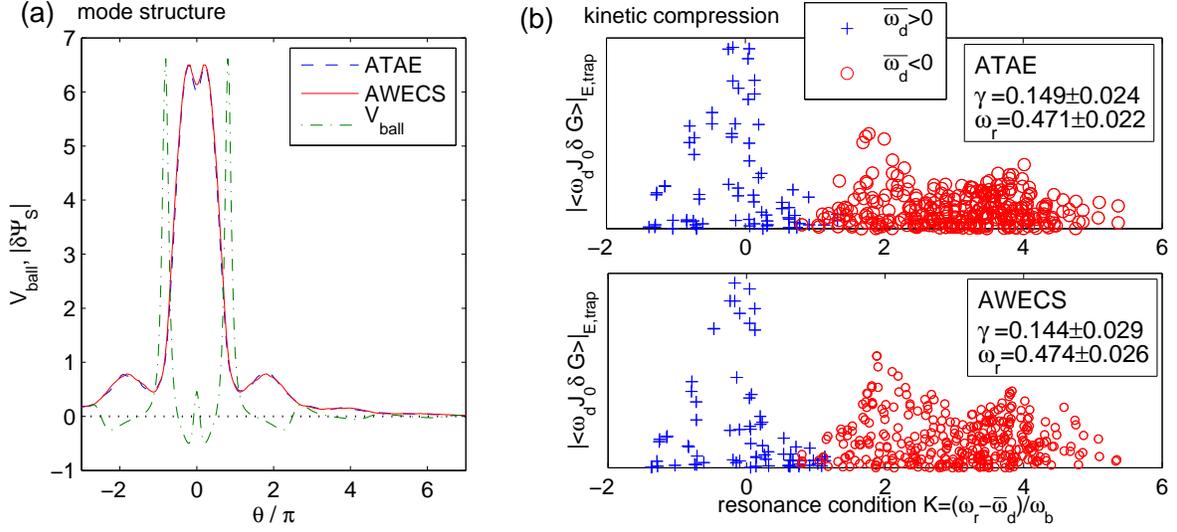}
\caption{Benchmark of \textsc{awecs} against \textsc{atae} for resonant excitation of $\alpha$TAEs by trapped energetic ions. In (a), the ballooning potential $V_{\rm ball}$ and the mode structure $|\delta\Psi_{\rm s}(\theta)| = \sqrt{f}|\dPsie(\theta)|$ are shown. In (b), the instantaneous contributions of individual phase-space markers to the kinetic compression term are plotted as a function of the resonance condition $K = (\omega_{\rm r} - \overline{\omega}_{\rm d}) / \omega_{\rm b}$, distinguishing between particles with positive ($+$) and negative (reversed) precessional drift ($\circ$). Most non-resonant particles are screened out in (b). The snapshot is taken at $t=400$ and the growth rates and frequencies are average values for $t \in [320,400]$. The errors represent the standard deviation due to discretization noise; the uncertainty of the mean values is smaller than this noise level by a factor $\sqrt{81} = 9$.}
\label{fig:bench_atae}%
\end{figure}

\section{Benchmark 3: Resonant excitation of $\bm\alpha$TAEs by energetic ions}
\label{sec:bench_atae}

The excitation of $\alpha$TAEs through trapped energetic ions was previously demonstrated by Hu \& Chen, 2005 (HC05) with the $\delta f$ PIC code \textsc{atae} \cite{Hu05}. In this section, we use \textsc{atae} results as a benchmark for \textsc{awecs}. \textsc{atae} employs a hybrid model consisting of an ideal-MHD core plasma (thermal electrons and ions) and a sparse population of trapped energetic ions. The dynamics of the latter is governed by the GKE (\ref{eq:model_vlasov_gke3}). The equations are sufficiently simple to be advanced with a leap-frog scheme, which allows to solve the GKE directly, without the need to apply the substitution $\delta G \rightarrow \delta g$, Eq.~(\ref{eq:model_final_gke_dGdg}).

In order to produce results comparable to those of \textsc{atae}, several adjustments needed to be carried out in \textsc{awecs}:
\begin{itemize}
\item  We omit all terms containing $\lambda'_s$; i.e., $\delta S_2$ and $\delta \Lambda_2$ and $\left<v_\parallel\lambda' J_1 \delta G\right>$, which are neglected in \textsc{atae}.

\item  For the GKE, we adopt the $\theta$-dependence of $\Omega_\kappa$, $\Omega_p$ and $\omega_{*s}$ as defined in \textsc{atae}, where these quantities are all $\propto 1/\hat{B}$.

\item  Let $k_\perp\rhoLi \ll 1$, take the ideal-MHD limit, $\delta E_\parallel = 0$, and approximate $\omega\dBp = -\Omega_p \delta\psi$. In \textsc{atae}, we have replaced
\begin{equation}
\Omega_p = -(\mu_0/B^2) {\bm k}_\perp \cdot (\hat{\bm b}\times\nabla P_{\rm c}), \nonumber
\end{equation}
where $P_{\rm c} = P - P_{\rm E}$, by
\begin{equation}
\Omega_p = -(\mu_0/B^2) {\bm k}_\perp \cdot (\hat{\bm b}\times\nabla P), \nonumber
\end{equation}
since the ordering $\beta_{\rm E}/\beta_{\rm c} \sim \O(\eps)$ used in Ref.~\cite{Hu05} does not apply for the parameters used.

\item  We eliminate effects of passing energetic ions. For trapped energetic ions, we retain only the ballooning term and the kinetic compression.
\end{itemize}

\noindent The most important modifications are the neglect of thermal ion FLR effects, the parallel electric field and passing energetic ions. While the former two are easily implemented by setting $\kiO \ll 1$ and $\dUe = 0$, the latter requires careful adjustments. After the parallel Amp\`{e}re's law is reduced to
\begin{equation}
\dEe \approx (1-\Gamma_{0{\rm i}}) \dPsie \approx b_{\rm i}\dPsie, \nonumber
\end{equation}

\noindent we are left with the vorticity equation
\begin{align}
f\kiO^2 \partial_t^2 \dPsie
=& \kiO^2 \left[f \dPsie'' + 2hh' \dPsie'\right]
+ \frac{\tauEi^n}{\tauei^T} \left<\omegad J_0 \delta G\right>_{\rm E,trap}
\label{eq:model_atae_de}
\\
&+ 4\omega_{*{\rm i}} \Omega_{\kappa{\rm i}}\dPsie
+ \kiO^2 \alpha_{\rm E,pass} g \dPsie
+ Z_{\rm E} \tauEi^n \tauiE^T \left<J_0^2\omegad\omega_*^T F_0\right>_{\rm E,trap} \dPsie. \nonumber
\end{align}

\noindent The terms on the second line of Eq.~(\ref{eq:model_atae_de}) are components of the so-called ballooning term. Their drift-kinetic limit may be summarized as $\kiO^2\alpha g\dPsie$ (not done here). The core component is
\begin{equation}
4\omega_{*{\rm i}} \Omega_{\kappa{\rm i}} = \kiO^2(\alpha - \alpha_{\rm E})g = \kiO^2\alpha_{\rm c}g, \nonumber
\end{equation}

\noindent the contribution of passing energetic ions is $\kiO^2 \alpha_{\rm E,pass} g$, and that of the trapped energetic ions is contained in the term $\left<J_0^2\omegad\omega_*^T F_0\right>_{\rm E,trap}$. The quantity $\alpha_{\rm E,pass}$ is calculated using the phase-space marker distribution for passing energetic ions at $t=0$. Since the velocity space moment $\left<...F_0\right>$ in Eq.~(\ref{eq:model_atae_de}) involves only trapped ions it needs to be evaluated numerically. Note that
\begin{equation}
f^{-1/2} (f \dPsie')' = \delta\Psi_{\rm s}'' - V_{\rm ball} \delta\Psi_{\rm s} - \alpha (g / f) \delta\Psi_{\rm s} \nonumber
\end{equation}

\noindent [cf.~Eq.~(\ref{eq:model_saw_eq})], where the last term, $\alpha (g/f)\delta\Psi_{\rm s}$, cancels with the drift-kinetic limit of the ballooning term.

In the benchmark simulation, the following parameters are used:
\begin{itemize}
\item  Physical parameters: $q=2.0$, $s=0.5$, $\alpha=2.1$, $\eps_{n{\rm c}}=\eps_{n_{\rm E}}=0.2$, $\vtE = 0.7$, $\kEO = 0.21$, $\tauei^T = 1.0$, $\eta_{\rm c} = 0.0$, $\eta_{\rm E} = 1.0$, $\eps = 0.2$, $Z_{\rm E} = M_{\rm E} = 1$. Some relevant parameters derived from the above are: $\tauEi^n = 5\times 10^{-2}$, $\tauEi^T = 392$, $\kiO = 0.010606$, $\beta_{\rm i} = 2.5\times 10^{-3}$, $\alpha_{\rm E} = 2.0$, $\alpha_{\rm E,trap} = 1.22$.

\item  Numerical parameters: $N_{\rm m} = 512\times 4\times 3$, $\theta_{\rm max} = 60$, $N_{\rm g} = 1024$, $\Delta t = 0.04$. Since \textsc{atae} uses a 2nd-order leap-frog scheme the time step adjusted to $\Delta t=0.02$.
\end{itemize}

\noindent This corresponds closely to the case shown in Fig.~1 of Ref.~\cite{Hu05}, with the difference that we use bounce angles in the range $\theta_{\rm b} \in [0.1^\circ,179.9^\circ]$.

The results are shown in Fig.~\ref{fig:bench_atae}. The mode structure in Fig.~\ref{fig:bench_atae}(a) shows an $\alpha$TAE with dominant (1,0) component. In Fig.~\ref{fig:bench_atae}(b), the instantaneous contributions of individual phase-space markers to the kinetic compression term are plotted as a function of the resonance condition $K = (\omega_{\rm r} - \overline{\omega}_{\rm d}) / \omega_{\rm b}$, with $\overline{\omega}_{\rm d}$ being the bounce-averaged magnetic drift frequency and $\omega_{\rm b}$ the bounce frequency \cite{Chen91}. Resonances can be observed at even values of $K$ which is consistent with the fact that $\alpha$TAE(1,0) is an eigenfunction with even parity. The precessional drift resonance, $K=0$, is due to particles with $\overline{\omega}_{\rm d} > 0$, whereas the drift-bounce resonances, $K \geq 2$, are due to particles with reversed drift $\overline{\omega}_{\rm d} < 0$.

The good quantitative ($\dPsie$, $\omega_{\rm r}$ and $\gamma$) and qualitative (wave-particle resonance) agreement between results obtained with \textsc{awecs} and \textsc{atae} shows that the interaction between trapped energetic ions and $\alpha$TAEs is accurately reproduced by \textsc{awecs} with the reduced vorticity equation (\ref{eq:model_atae_de}). Our preliminary studies indicate that the inclusion of thermal ion FLR effects, $\delta E_\parallel \neq 0$ and passing energetic ions could significantly modify the results. To our knowledge, such a case has not been studied before and \textsc{awecs} will be used to advance into this area.

\begin{figure}[tbp]
\centering
\includegraphics[width=0.84\textwidth]
{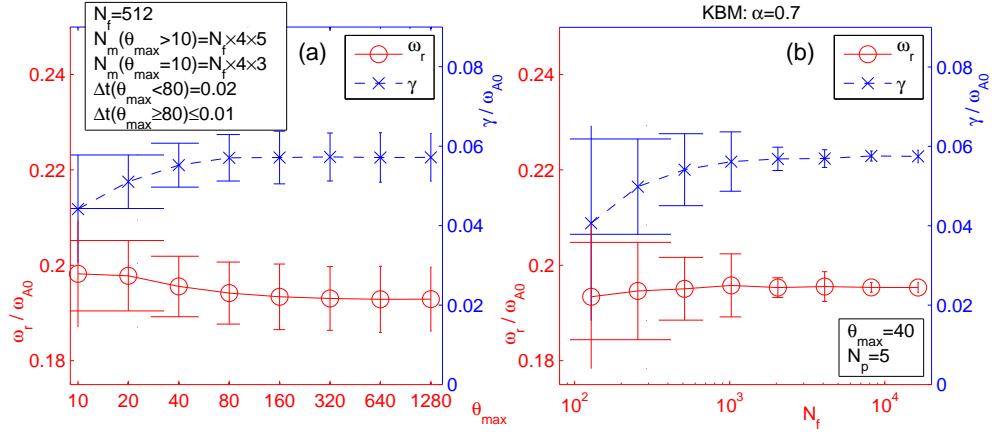}
\caption{Convergence study for a case from Fig.~\protect\ref{fig:bench_saw_zhao} with $\alpha = 0.7$ and $\eta_{\rm i} = 0$; i.e., a KBM without ITG-drive. (a) shows convergence with the simulation domain size $\theta_{\rm max}$, and (b) convergence with the number of markers $N_{\rm m} = N_{\rm f,ipass}\times 4\times N_{\rm p}$ (controlled by $N_{\rm f,ipass}$).}
\label{fig:accurate_conv_zhao}%
\end{figure}

\begin{figure}[tbp]
\centering
\includegraphics[width=0.84\textwidth]
{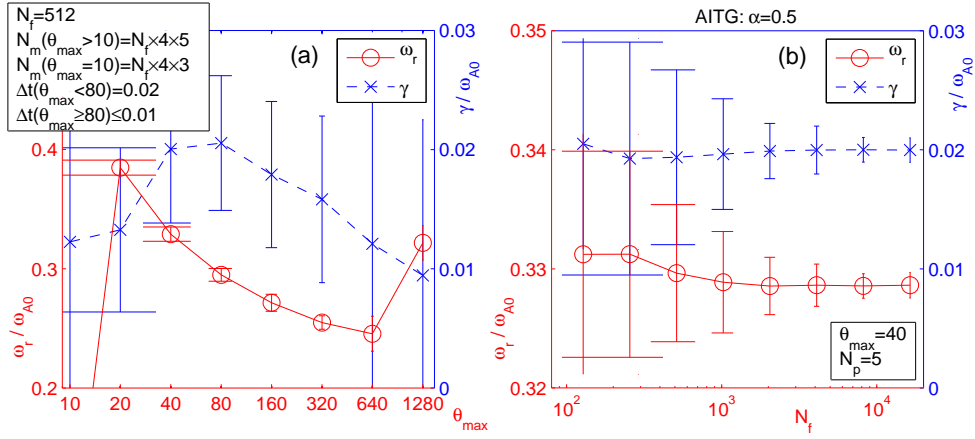}
\caption{Convergence study for a case from Fig.~\protect\ref{fig:bench_saw_zhao} with $\alpha = 0.5$ and $\eta_{\rm i} = 2.5$; i.e., at the instability threshold of AITG modes. (a) shows that convergence with the simulation domain size $\theta_{\rm max}$ is not achieved. In (b), convergence with the number of markers $N_{\rm m} = N_{\rm f,ipass}\times 4\times N_{\rm p}$ (controlled by $N_{\rm f,ipass}$) is demonstrated for a given domain size $\theta_{\rm max} = 40$.}
\label{fig:accurate_conv_zhao_margin}%
\end{figure}

\begin{figure}[tbp]
\centering
\includegraphics[width=0.84\textwidth]
{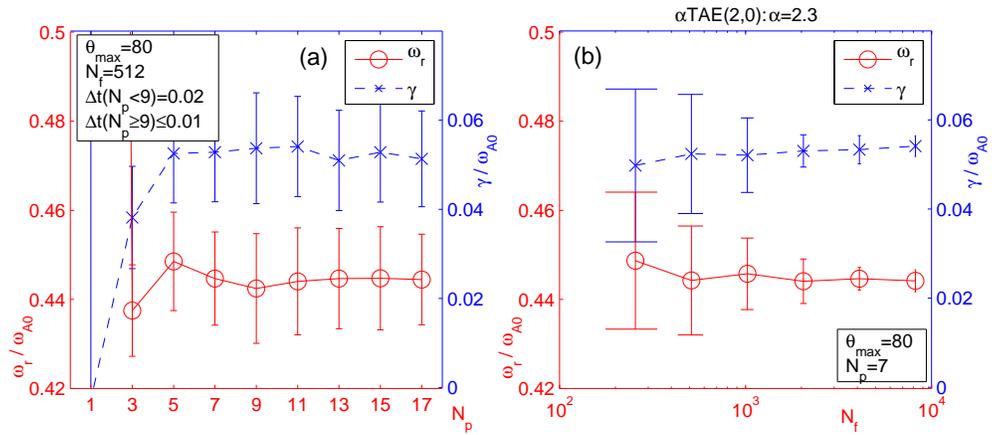}
\caption{Convergence study for a case from Fig.~\protect\ref{fig:bench_saw_hirose} with $\alpha = 2.3$ and $\eta_{\rm i} = 2.0$; i.e., a predominantly 2nd-order $\alpha$TAE driven by ITG. (a) shows convergence with the number of periods $N_{\rm p}$ in which markers are loaded, and (b) convergence with the number of markers $N_{\rm m} = N_{\rm f,ipass}\times 4\times N_{\rm p}$ (controlled by $N_{\rm f,ipass}$).}
\label{fig:accurate_conv_hirose}%
\end{figure}

\section{Numerical convergence}
\label{sec:accurate_converge}

Examples for numerical convergence with the domain size $\theta_{\rm max}$, the number of markers $N_{\rm m} = N_{\rm f} \times 4 \times N_{\rm p}$, and the number of marker loading periods $N_{\rm p}$ are shown in Figs.~\ref{fig:accurate_conv_zhao}, \ref{fig:accurate_conv_zhao_margin} and \ref{fig:accurate_conv_hirose}. The error bars indicate the level of discretization noise, which decreases with increasing number of markers. Only cases with $\eps=0$ and without energetic ions are considered here; i.e., $N_{\rm p} = N_{\rm p,ipass}$ (passing thermal ions only).

In Fig.~\ref{fig:accurate_conv_zhao}, a convergence study is shown for a KBM in the case considered by Dong \textit{et al.}, 2004 (DCZJ04) \cite{Dong04} and Zhao \& Chen, 2002 (ZC02) \cite{Zhao02} (cf.~Fig.~\ref{fig:bench_saw_zhao}). DCZJ04 used $\theta_{\rm max} \approx 23$ and ZC02 used $\theta_{\rm max} \leq 50...100$, while Fig.~\ref{fig:accurate_conv_zhao}(a) indicates that at least $\theta_{\rm max} \sim 40$ is required in \textsc{awecs} for numerical convergence. It is thus possible that the results by Dong~\textit{et al.} \cite{Dong99, Dong04} are not fully converged with respect to the simulation domain size. The \textsc{awecs} results in Fig.~\ref{fig:bench_saw_zhao} were obtained with $N_{\rm m} = 512\times 4\times 5$ markers, and the equivalent number of markers used in ZC02's calculations is $N_{\rm m} \approx (315...630)\times 4\times 5$, which appears to be sufficient according to Fig.~\ref{fig:accurate_conv_zhao}(b).

Near the stability boundary, convergence with $\theta_{\rm max}$ becomes problematic, as is illustrated in Fig.~\ref{fig:accurate_conv_zhao_margin}(a) for an AITG mode \cite{Zhao02} (cf.~Fig.~\ref{fig:bench_saw_zhao}). This result shows that it may be difficult to obtain an accurate value for $\acriti$ (and possibly $\acritii$, as well) other than through extrapolation from converged results away from the stability boundary. Despite the lack of convergence with respect to $\theta_{\rm max}$, the results still converge well with the number of markers for a given $\theta_{\rm max}$, as can be seen in Fig.~\ref{fig:accurate_conv_zhao_margin}(b).

In Fig.~\ref{fig:accurate_conv_hirose}, a convergence study is shown for higher-order $\alpha$TAEs (with dominant 2nd-order mode) in the 2nd stable domain in a case considered by Hirose \textit{et al.}, 1994 (HZE94) \cite{Hirose94}. Due to the broad mode structure, markers must be loaded in at least $N_{\rm p} = 7$ periods, as was done in Fig.~\ref{fig:bench_saw_hirose} from which this case was taken. The number of markers was $N_{\rm m} = 512\times 4\times 7$, which is also found to be sufficient.

Given the limitations of the model used, the accuracy of the results shown in Figs.~\ref{fig:bench_saw_zhao}--\ref{fig:bench_saw_hirose} may be considered to be sufficient to delineate the qualitative features of the modes studied. For this purpose, it is usually not necessary to carry out the expensive calculations required for full convergence.

\section{Conclusions and discussions}
\label{sec:conclusion}

A 1-D linear gyrokinetic code called \textsc{awecs} has been developed to study the kinetic excitation of Alfv\'{e}nic instabilities in a high-$\beta$ tokamak plasma. The model equations and the numerical scheme are described and the code has been tested carefully. In particular, it is shown that \textsc{awecs} reproduces successfully essential properties of ESITG and AITG modes, KBMs, and $\alpha$TAEs. Benchmarks against results in Refs.~\cite{Dong92, Zhao02, Idomura04, Hirose94, Hu05} are regarded as successful. However, discrepancies persist in comparisons with Refs.~\cite{Dong99, Dong04}.

While the real frequencies calculated by \textsc{awecs} have also been confirmed by MHD shooting code calculations, the quantitative accuracy of the growth rates is more difficult to show, and it is here where the main discrepancy with Refs.~\cite{Dong99, Dong04} lies. The code used by Zhao \& Chen \cite{Zhao02} is most directly comparable to \textsc{awecs} (without energetic ions), and here the growth rates agree well. Numerical convergence studies in several typical cases indicate that accurate calculations require a larger simulation domain than used in Refs.~\cite{Dong99, Dong04}. Furthermore, outgoing boundary conditions are required to accurately reproduce properties of modes subject to continuum damping, such as $\alpha$TAEs. Hence, the lack of numerical convergence and sensitivity to the boundary conditions may explain the discrepancies, as was previously suggested in Ref.~\cite{Zhao02}.

Overall, it can be concluded that \textsc{awecs} is functioning properly and that it can be used for further research. Since benchmarking cannot rule out all possible modeling and programming errors, \textsc{awecs} will undergo continuing scrutiny while in operation. Code maintenance and extensions are simplified by the modular structure of the code and the application of principles of object-oriented programming. Interactive graphical tools were developed using \textsc{matlab} in order to assist the user in data analysis and post-processing tasks.

The demonstration of $\alpha$TAE excitation in the cases studied by Hirose \textit{et al.}, 1994 \cite{Hirose94} and Dong \textit{et al.}, 2004 \cite{Dong04} constitutes the first successful application of this code. Note that, in these earlier works, the observed instabilities were not identified as $\alpha$TAEs, which were discovered more recently \cite{Hu04}. Details about the physics of $\alpha$TAE excitation through wave-particle interactions with thermal and energetic ions will be reported elsewhere. With the use of \textsc{awecs}, many other interesting problems may be addressed in the future, including the physics of BAEs, TAEs and EPMs.

In this paper, the CHT $s$-$\alpha$ model equilibrium is adopted in order to be able to carry out comparisons with previous studies, and, thereby, benchmark the code. Currently, work is underway to implement the local equilibrium model developed by Miller \textit{et al.} \cite{Miller98} which will allow us to explore the high-$\beta$ regime more relevant to experiments. Further extensions are under consideration, such as the inclusion of effects due to magnetically trapped electrons.

Finally, two cautionary notes remain to be added. First, since the CHT $s$-$\alpha$ model is derived under the assumption that $B$ is independent of $\theta$, there is a certain degree of arbitrariness in the way how the extension to a model with finite aspect ratio, $\eps > 0$ and, thus, variable $B(\theta)$, is carried out. Second, all velocity integrals not arising from the transformation $\delta G \rightarrow \delta g$ must be evaluated analytically or using phase-space markers distributed uniformly along $\theta$. On the other hand, those integrals arising from the transformation $\delta G \rightarrow \delta g$ must be evaluated numerically, utilizing the same markers used to calculate $\delta g$ (nonuniform in $\theta$). If this is not done, inconsistencies arise due to a mixing of $\theta$-independent and $\theta$-dependent densities. Preliminary tests where the ballooning term contains $\hat{B}$, so that the cancellation with the $\alpha g/f$ term arising from the substitution $\dPsie\rightarrow \delta\Psi_{\rm s}$  [cf.~Eq.~(\ref{eq:model_saw_cold}) and the note following Eq.~(\ref{eq:model_atae_de})] is incomplete, have shown that such inconsistencies may modify the ballooning potential to such a degree that unstable modes appear in the second MHD ballooning stable domain even without wave-particle interactions. In summary, care must be exercised when the CHT $s$-$\alpha$ model employed in this paper is used with finite aspect ratio, $\eps > 0$.

\section*{Acknowledgments}

One of the authors (A.B.) would like to thank Z.~Lin and Y.~Nishimura (UCI), as well as F.~Zonca (Euratom-ENEA, C.R. Frascati) for helpful discussions. The modular structure of \textsc{awecs} was partly inspired by a PIC framework developed at UCLA by V.~Decyk and C.~Norton. Further ideas, in particular with regard to parallelization using MPI, were adopted from the code \textsc{d3d} by B.~Scott (Max-Planck-IPP Garching). This research is supported by U.S. DoE Grant DE-AC02-CH0-3073, NSF Grant ATM-0335279, and in part by SciDAC GSEP.

\section*{Erratum}

In the published version of this work [A.~Bierwage \& L.~Chen, \textit{Communication of Computational Physics} \textbf{4}, 457 (2008)], the following errors were found:
\begin{enumerate}
\item  The coefficient in front of $\dUe$ on the left-hand side of Eq.~(2.33) was $\left(1/\tauei^T + \Gamma_{0{\rm i}} - H_\omega\right)$ but should read
\begin{equation}
\left[1/\tauei^T + \Gamma_{0{\rm i}} - (H_\omega - Z_{\rm E}\tauEi^n\tauiE^T)\right].
\end{equation}

\item  The coefficient in front of $\dCe$ on the left-hand side of Eq.~(2.36) was $(\hat{B}^2 + A_\Sigma^2A_\omega)$ but should read
\begin{equation}
(\hat{B}^2 + \Ti A_\Sigma^2A_\omega).
\end{equation}

\item  The title of Section 4.2 was ``Zero aspect ratio'' but should read ``Zero inverse aspect ratio.''
\end{enumerate} 

\clearpage




\end{document}